\documentclass{article}
\usepackage[dvipsnames]{xcolor}
\usepackage[utf8]{inputenc}
\usepackage{changepage}
\usepackage{dsfont}
\usepackage[english]{babel}
\usepackage{times}
\usepackage[T1]{fontenc}
\usepackage[official,right]{eurosym}
\usepackage{url}
\usepackage{eso-pic}
\usepackage{tabularx}
\usepackage{rotating}
\usepackage{amsmath}
\usepackage{lscape}
\usepackage{tikz}
\def\checkmark{\tikz\fill[scale=0.4](0,.35) -- (.25,0) -- (1,.7) -- (.25,.15) -- cycle;}
\usepackage{multirow}
\usepackage{natbib}
\usepackage{graphicx}
\usepackage{tikz}
\usetikzlibrary{arrows,automata}
\usepackage{pgf}
\usepackage{pgfplots}
\usepackage{amssymb}
\usepackage{mathtools,zref-savepos}
\usepackage{version}
\usepackage{float}
\usepackage{comment}
\usepackage{setspace}
\usepackage[top=4cm, bottom=4cm, left=4cm, right=4cm]{geometry}

\def\qed{\hbox to\hsize{\hfill\vrule height 1.6ex width 1.5ex depth -.1ex}}

\title{Is the Capability approach a useful tool for decision aiding in public policy making?}
\author{Nicolas Fayard$^*$, Chabane Mazri$^\dag$, Alexis Tsoukiàs$^*$ \\
($^*$)LAMSADE-CNRS, PSL, Université Paris-Dauphine, ($^\dag$)INERIS }
\date{}
\begin{document}
\thispagestyle{empty}

\enlargethispage*{8cm}
 \vspace*{-38mm}

\AddToShipoutPictureBG*{\includegraphics[width=\paperwidth,height=\paperheight]{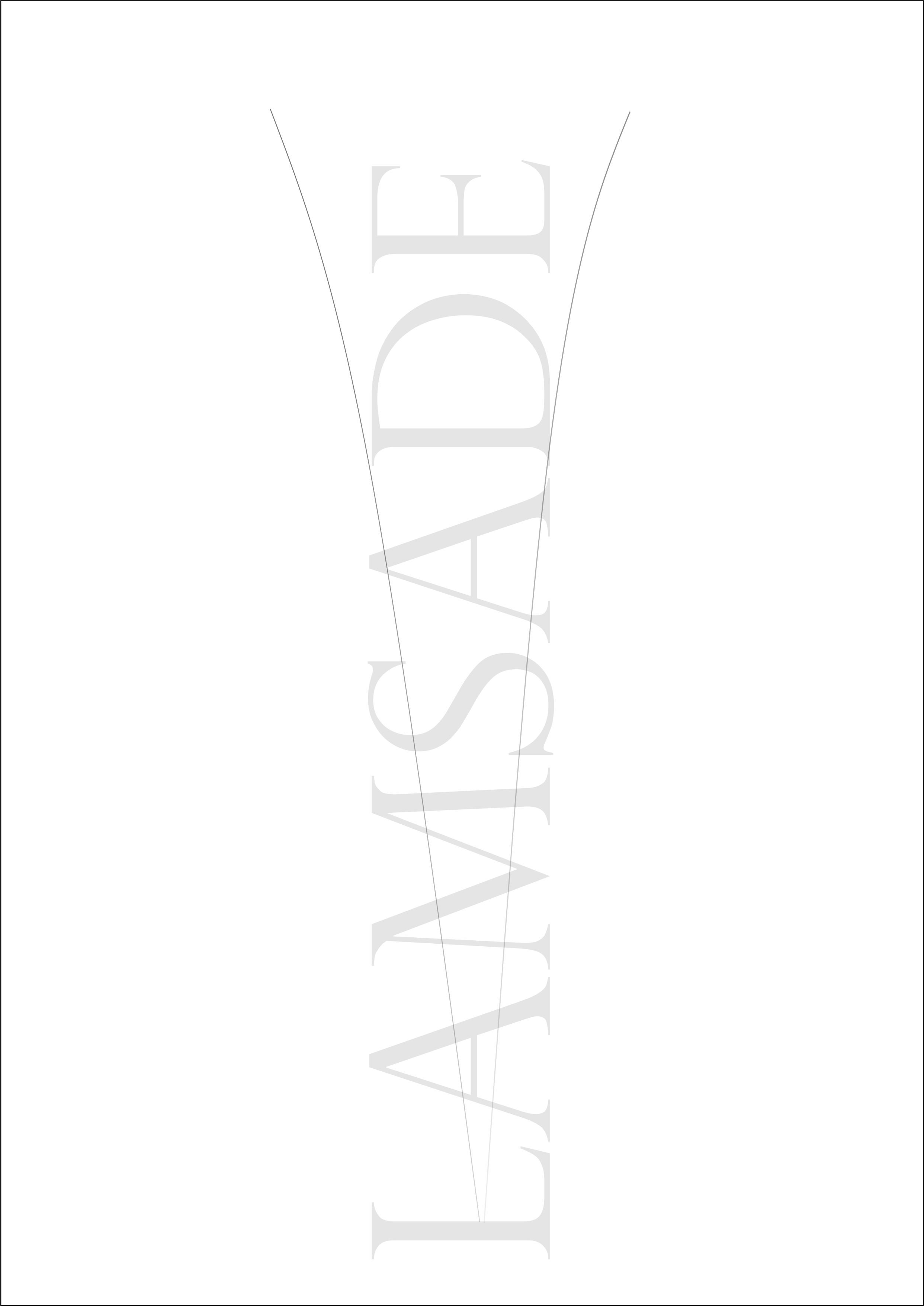}}

\begin{minipage}{24cm}
 \hspace*{-28mm}
\begin{picture}(500,700)\thicklines
 \put(60,670){\makebox(0,0){\scalebox{0.7}{\includegraphics{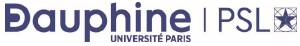}}}}
 \put(60,70){\makebox(0,0){\scalebox{0.3}{\includegraphics{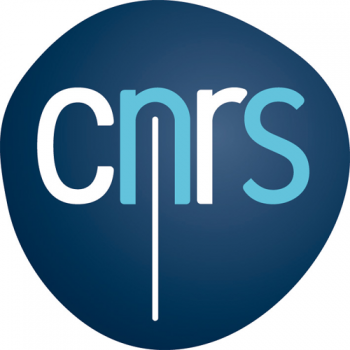}}}}
 \put(320,350){\makebox(0,0){\Huge{CAHIER DU \textcolor{BurntOrange}{LAMSADE}}}}
 \put(140,10){\textcolor{BurntOrange}{\line(0,1){680}}}
 \put(190,330){\line(1,0){263}}
 \put(320,310){\makebox(0,0){\Huge{\emph{398}}}}
 \put(320,290){\makebox(0,0){January 2021}}
 \put(320,210){\makebox(0,0){\Large{Is the Capability Approach a useful}}}
 \put(320,190){\makebox(0,0){\Large{tool for decision aiding in public policy making?}}}
 \put(320,100){\makebox(0,0){\Large{Nicolas Fayard, Chabane Mazri, Alexis Tsoukiàs}}}
 \put(320,670){\makebox(0,0){\Large{\emph{Laboratoire d'Analyses et Mod\'elisation}}}}
 \put(320,650){\makebox(0,0){\Large{\emph{de Syst\`emes pour l'Aide \`a la D\'ecision}}}}
 \put(320,630){\makebox(0,0){\Large{\emph{UMR 7243}}}}
\end{picture}
\end{minipage}

\newpage

\addtocounter{page}{-1}

\maketitle

\begin{abstract}
This paper aims at proposing a model representing individuals' welfare using Sen's capability approach (CA). It is the first step of an attempt to measure the negative impact caused by the damage at a Common on a given population's welfare, and widely speaking, a first step into \emph{modelling collective threat}.

The CA is a \emph{multidimensional} representation of persons' well-beings which account for \emph{human diversity}. It has received substantial attention from scholars from different disciplines such as philosophy, economics and social scientist. Nevertheless, there is no empirical work that really fits the theoretical framework.

Our goal is to show that the capability approach can be very useful for decision aiding, especially if we fill the gap between the theory and the empirical work; thus
we will propose a framework that is both usable and a close representation of
what capability is.

\end{abstract}

%\tableofcontents

\section{Introduction}

To a large extent aiding to design public policies consists in introducing elements of rationality (under different forms) within a public decision process. Such elements come under different forms of evidence and argumentation. A typical example of such rationalisation is ``Cost Benefit Analysis''(CBA, \cite{DasguptaPearce72}): in reality nobody considers the result of a CBA to have a normative validity, but almost everybody is ready to accept it as a common ground for different stakeholders discussing the interest and acceptability of undergoing a certain project. Actually it is practically considered \textbf{THE} legitimating exercice for almost any institutional decision process concerning large public investments.

%The CA aims to represent individuals' well-being as the
%freedom they have to choose to live a life that they valued, or their real
%opportunity of life. It is an normative theoretical framework that is an
%alternative to welfare economics and other theory of justice. The capability
%approach is framework aims to represent individuals' well-being as the freedom
%they have to choose to live a life that they valued, or their real opportunity
%of life. It is an normative theoretical framework that is an alternative to
%welfare economics and other theory of justice.

``Rationally speaking'' designing a policy which is expected to have an impact
upon the citizens' welfare implies being able to: \\
 - observe the present situation and distribution of welfare; \\
 - anticipate the impact of doing nothing; \\
 - anticipate the impact of implementing a policy.

In other terms if a policy is expected to have any impact upon the citizens'
welfare it makes sense to try to measure it: as it stands presently and as it
could stand under different possible scenarios. However, welfare is a complex
issue, implying multiple dimensions and aspects, impacting and being perceived differently among different segments of the society, being distributed unequally among the citizens. Moreover, measuring welfare is itself a policy, since any measurement will need to make choices about what and how to measure. Under such a perspective it is unlike that a single figure welfare measurement can be of any utility for effective policy design purposes.

Further on, welfare appears to be very much related on how citizens perceive
themselves as being appropriately endowed and how much they feel free to use
their commodities in order to realise their own aspirations. From that point of view, welfare appears to have a strong subjective dimension: as a consequence any attempt of ``measuring'' welfare needs to be able to capture this special feature.

Besides the above two remarks we need to consider a third difficulty: part of
the ``endowments'', allowing citizens to live as they do, are ``commons'',
goods which are shared with other citizens, but consumed individually. How such ``commons'' affect each citizen's welfare and how much this welfare could be reduced in case any of these ``commons'' gets lost (totally or partially)? The contribution of ``fresh water availability'' to a community's citizen welfare is far beyond the price for each liter of water consumed by each citizen.

Let's resume. We need to be able to ``measure'' welfare and the impact policies can have upon ``welfare'', but such measurement needs to consider: \\
 - the multidimensional nature of welfare; \\
 - the subjective dimension of welfare; \\
 - the impact of the commons upon welfare; \\
 - the different impact a policy or an event can have upon different groups of citizens.

The above constraints represent a challenge for decision analysts: if our tools are aimed to help (among others) policy makers to design policies and to improve how policies are designed we need to provide appropriate methods taking into consideration the above discussion. This becomes particularly relevant when we consider policies which are expected to be a response against potential threats to communities, territories and commons: what we call collective threats.

How to evaluate the negative impact upon welfare after the damage suffered by a good shared by multiple individuals? It is clear that a financial analysis is not sufficient to represent this impact. The damage of a road cannot be reduced to its economic cost, the real negative impact on individuals is their incapacity to use it anymore, which can lead to difficulties in reaching facilities such as schools, workplaces, hospitals ...

Our attempt is to propose a framework that supports public decisions processes occurring in a \emph{policy cycle}, within a \emph{Policy Analytics} framework \citep{tsoukias2013policy,de2016evidence, daniell2016policy}. This concept aims at supporting policy makers in a way that is meaningful, operational and legitimate, by developing, in particular, methods that take into account values of different stakeholders. However, we need a theory about how to consider welfare. For this purpose, in this paper we explore the Capability Approach (CA) \citep{sen1979equality,sen1993capability,sen1985commodities,
sen1999development, sen2009idea} as a framework allowing to develop appropriate decision aiding for public policy design. Our proposal is that CA could offer a common ground to different stakeholders assessing the impact of a policy to the citizens' welfare, offering a certain number of advantages with respect to other approaches aiming at measuring welfare (although technically more complicated and certainly more expensive to conduct in terms of analysis).

The CA aims at representing what an individual is effectively capable to do and to be. It is thus, a good framework to use in our context. However, this theory suffers from a lack of empirical development, and is generally used as a multidimensional framework to evaluate individual actual life, rather than an evaluative framework over the set of lives that they are capable to achieve.

The first section of this report is about Capability. It begins with an
introduction of welfare economics and social choice theory, which aims to
develop techniques to evaluate potential public policies based on their impact on people well-beings. Then the Rawlsian theory of justice will be presented. It can be interpreted as a resource-based approach of well-being that has greatly influenced Sen's theory. Next, we will see a criticism of those two approaches leading to the capability theory, presented and discussed later. Finally, we will focus on the empirical use of this framework.

In the second section, our problem of modelling collective threat will be
introduced. We will then see the interest of the capability approach in this
evaluation exercise. After that, we will present our framework and conclude.

\section{Welfare economics and capability approach}

Sen's Capability Approach is a theoretical framework which concerns the
identification of individuals' well-being, as well as how to consider equality and justice. It focuses on the individual freedom to choose to live a life that one values.
%This conception is increasingly influential in different academic fields.
Sen has been influenced by different thinkers such as Aristotle, Marx, Smith
and Rawls, and also by the welfare economics and social choice theory. To have a good comprehension of Sen's approach, we will first introduce the fundamental aspects of these theories, and then we will introduce capability approach (CA).
%Following Sen,``welfarism
%%is the name given to the normative approaches that rank social states on the \emph{sole} basis of
%the distribution of welfare levels achieved by individuals in those states.''

\subsection{Welfare economics and Rawls's theory}

Welfare economics is a branch of economics which aims to evaluate ``social welfare'' defined as the representation of the ``goodness'' of the social state \cite{sen1991welfare}. This is not a complete description
of the whole welfare economics, but rather a presentation of the necessary
background for a better understanding of the influences of the capability
approach. Welfare economics has grown through different developments, therefor which we present following an historical order.

\subsubsection{The hold welfare }\label{hold}

It can be said that the roots of welfare economics is utilitarianism
\citep{hicks1939foundations}. It is a consequentialist ethical theory
\citep{sinnott2003consequentialism}, meaning it considers an action being
good or bad, based on its consequences. For utilitarians, the utility
is the representation of welfare, or the happiness of an individual. In
\cite{bentham1789introduction}, the founder of utilitarianism, he describes the
utility as following:

\begin{adjustwidth}{30pt}{30pt}
\emph{``Nature has placed mankind under the governance of two foreign matter,
\textbf{pain} and \textbf{pleasure}. It is for them alone to point out what we
ought to do, as well as what we should do''}
\end{adjustwidth}

Bentham was a hedonist, for him, the utility is the difference between the
\emph{pain} and the \emph{pleasure}, and everyone has to act in order to
maximize the pleasure and minimize the pain. This principle of utility should
be considered as the foundation of our moral judgement and political decision.
Nothing is moral on its own, we don't have to act to maximize liberty or
justice, but we have to judge actions based on their consequences over the
utility. An action is considered as good, if it implies \emph{the greatest
happiness of the greatest number} \citep{bentham1789introduction}.

For Bentham, the utilitarian approach is descriptive and prescriptive. It is
descriptive because it explains why people act as they do. They are just
maximising their utility, or their happiness. It is also prescriptive because
it allows us to identify if an action is morally good or bad, by looking if it maximises the utility of the greatest number.

Bentham's student, John Stuart Mill, extended this work, introducing the
notion of higher and lower pleasure \citep{mill1863utilitarianism}. When all
forms of pleasure are equals for Bentham, the intellectual and moral pleasure
are higher than the physical one's for Mill. Then, some moral judgement has to be done in order to define what is a higher pleasure.

Despite the intuitive appealing of utilitarianism, there are some critical aspects (\cite{Robbins1932}): judging an alternative only on the utility of the greatest number, summing the utility of the individuals, utilitarians don't take into account distributional aspects. Considering two individuals, they will prefer an alternative of the form: (person 1's utility: 99, person 2's utility: 1), to an alternative of the form (person 1's utility: 49, person 2's utility: 49), even if the second alternative seems fairer. Utilitarianism also need to do two fundamental assumptions;

\begin{itemize}
    \item Utility is \emph{cardinal} (it can be represented by a quantity
        upon an interval scale).
    \item Utility allows \emph{interpersonal comparison} (we can compare
        utility of individuals, and add them).
\end{itemize}

It means that we are able to measure the difference of value between any two
states, and this measurement is commensurable among individuals, implying that we are able to compare the difference among values of a first individual with the difference among values of a second one.

%\begin{note}
 % Cardinal is a notion implying more than a simple numerical representation.
  %Implies ``quantities'' (ratios, distances etc.). This needs to be explained.
  %More precisely utility means we are able to measure the difference of value
  %between any two states and that this measurement is commensurable among
  %individuals which means that I am able to compare the differences among
  %values of individual X to the differences of values of individual Y. Under
  %such a perspective commensurability does not imply additivity. You need
  %further axioms to get that. See Savage's framework.
%\end{note}

%\paragraph{Cost-Benefit Analysis (CBA)}
{\em Cost-Benefit Analysis (CBA)} is a set of techniques well-spread in decision making having deep roots with Welfare economics  \citep{pearce1983origins}, and usually apply in the evaluation of public sector projects (economics, transport, health, environment...). It is based on the idea that decisions should be evaluated in terms of their consequences, a project should be chosen if its benefit are greater than its cost.

Let's give a small an overview of the hearth of the CBA. Citizens are viewed as ``consumers'' and their marginal utility are derived from their consumption of goods. As it is assumed that stakeholders are living in a perfect market economy and are maximizing their utility, we can find a ``marginal utility income rate'', \emph{e.i. utility can be derived from income}, see \cite{boadway1974welfare}; \cite{johansson1991introduction}, Chapter 9; and \cite{bouyssou2000evaluation}, Chapter 5. Such a perfect market does not exist for all goods and services (think of public good such as education, health or Commons such as forests or bridges), their shadow prices have to be determined. In the same way, the price of externalities (such as noise, or the beauty of an installation) or the price of a life have to be evaluated.

A project is seen as a \emph{change in the net supplies of commodities} \citep{dreze1987theory} and should be selected if its benefit exceed its cost over the period of time to consider \emph{e.i. if the social Welfare increases over the select period}. As the utility of income is supposed being greater today than tomorrow, a discount rate is applied. Finally, note that the prices of goods are considered being fixed, then, a CBA can only be applied on projects that \emph{marginally} change the current state, \emph{e.i. small project infinitesimally perturbing the market price}.

\emph{Cost-Benefit Analysis} is a typical example of how utilitarianism is practically used in terms of decision aiding. Consider a project which is expected to have an impact upon the life and welfare of the citizens of a given territory. Given the expected consequences of the project and considering the citizens as individual consumers the utility of implementing the project is the sum of utilities of each single citizen/consumer (NB: each citizen has exactly the same utility!). Then in order to compute such utility we consider that each potential consequence is measurable at a real/proxy market revealing the citizens/consumers preferences. In other terms the ``prices'' of such consequences, observable directly or indirectly through the markets, represent the utilities lost or perceived by each citizen/consumer and we only need to sum them.

\vspace{5mm}

These assumptions (about interpersonally comparable utility) have been criticized by \cite{Robbins1932,robbins1938interpersonal}: utility cannot be objectively compared because it is derived from mental states. One cannot measure the utility of someone else, which makes interpersonal comparisons impossible. This led to the revision of how we should judge an action. Keeping the notion of utility, the ``new welfare economics'' (emerged during the 1930s) reject the sum of individuals' utilities as a basis for the judgement of an action.

%\subsubsection{Utilitarism (Jeremy Bentham)}
%Consequentialist theory. Utility is cardinal and can be compared between
%different people. Two situations are indifferent if the sum of the utilities of
%the individuals are equal (even if the social state is unfair).

\subsubsection{New welfare economics}

%What we can consider as the new welfare economics emerge from the critics the
%classical utilitarianism, and more specifically, about the carnality and the
%interpersonal comparison of utilities.

%\subsubsection{Pareto optimality (Vilfredo Pareto) and The two fundamental theorems of welfare
%economics. (Kenneth Arrow, Gerard Debreu)}

The new welfarists' uncontroversial argument is that social states should be
judged according to their utilities, but without any comparison between
utilities of different individuals. The new foundations had to use the Pareto principle (see \cite{Pareto06}).

\paragraph{Pareto principle}\label{hold}

\emph{Pareto principle: } A social state is said to be Pareto optimal, if and
only if we cannot increase the utility of an individual without decreasing the utility of another one. However, the use of this principle is far from being sufficient to reply to the critiques (\cite{sen2018collective}).

Not allowing any comparison between different individuals' utilities is a very restrictive informational basis to evaluate alternatives. New welfarism uses Pareto optimality for this purpose, or in other words, the model chosen for social improvement is the Pareto principle. Using preferences over
alternatives\footnote{In the case of complete non-interpersonal comparison,
considering cardinal or ordinal utilities leads to the same results}, we can
consider that the Pareto principle is an unanimous preference ranking, because if one individual is worst off with a new situation, we should not move to this alternative. We can improve a situation only if it is not damaging others.

Note that the fact that cost-benefit analysis (presented in the section \ref{hold}) is meeting the Pareto principle is questionable.
One idea from \cite{harberger1971three} is that if benefits are greater than costs, winners can compensate losers more than their losses, while still guaranteeing to always win something, generating a Pareto improvement.
This argument is not necessarily valid in all cases (as shown in \cite{boadway1974welfare}) and whether CBA is a new welfarist approach or not is not the purpose of the paper. We simply consider CBA as an utilitarian approach since it is ultimately a test on sum of individual utilities.

With the Pareto principle, it is very difficult to find a social improvement in real life. Indeed, public policies will most of the time hurt the utility of some individuals, even if these policies seem to be morally positive. For
example, let's consider two alternatives where everyone has the same
preferences over commodities. In the first situation $(a)$, commodities are
well distributed, so everybody enjoys a good level of utility. In a second
situation $(b)$, all commodities are given to an individual, and the others are left with nothing. Here, the Pareto criteria does not help us to compare $(a)$ and $(b)$, because none of them Pareto dominates the other one, as for the classical utilitarian approach, the Pareto principle is not concerned with the distribution of utility. However, intuitively $(b)$ seems to be less desirable than $(a)$.

Improvement by Pareto comparison is in practice very unrealistic if the purpose is to design public policies, because these will typically benefit to some individuals at the cost for others. The further reduction of informational basis, by removing interpersonal comparison, leads to the incomparability of most of the alternatives. Then, some new criterion has to be taken into account. Some judgements over alternatives have to be done; for this purpose Abram \cite{bergson1938reformulation} developed the concept of Social welfare function.

\paragraph{Social Welfare Function}

The critic made by Robin over interpersonal comparison, was that they are not
objective. As argued in \cite{weymark2016social}, Robbins didn't claim that
interpersonal comparison couldn't be done, or that they were not usually made, his point was that utilities are normative, and then cannot be
\emph{objectively} determined.  One way to respond to this criticism has been
to introduce Bergson-Samuelson social welfare function
\citep{kaushik2011functionings}. This concept emphasises that the selection of a ``best'' solution from the set of Pareto efficient one's, should be based on ethics. It is necessary to add more criteria to be able to judge different policies that are Pareto efficient, and these criteria should be ``ethical''. The role of scientists is then to analyze the outcome of different values of judgement, but the values of judgement, coming from someone else.

First, we have to calculate the utility of individuals for each alternatives of $A$, using an individualistic utility function
\citep[Chapter~3]{sen2018collective}. This function takes into account relevant elements such as; how much of a good is consumed by an individual, what is his/her amount of work, or what are the characteristics of the environment. From those utility functions, we obtain an ordering on $A$ for each individual. Then, we have to apply a social welfare ordering, which is going to order the alternatives according to the individuals' utilities, and \emph{values of judgements} that we want to use.

The new welfare foundations use ordinal and non-interpersonal utilities as
informational bases. This leads us to social choice theory: how to establish the preferences of a society given the preferences of its members.

\paragraph{Social choice theory}

We can claim that the theoretical work and methods of social choice theory started with \cite{marquis1785essai} (mathematician) and
\cite{borda1784memoire} (engineer). They were concerned with the development of a framework that would allow rational and democratic decisions for a given
group of citizens. They both developed voting procedures, that consist into
aggregating individual rankings over alternatives in a unique \emph{social
ranking}. They both obtained frameworks containing weaknesses that we do not discuss here (see instead: \cite{black1958theory, farquharson1969theory, sen2018collective}). Yet, social choice theory really started being developed thanks to Arrow's works.

\paragraph{Arrovian Social Welfare Functions}

As previously noted, the new welfare economics and the social choice theory
uses the same informational base. By working on Social Welfare Functions (SWF), Arrow made deep changes in both theories. In \cite{arrow1951social}, he introduced his \emph{General Possibility Theorem} (also known as the
\emph{Arrovian impossibility theorem}), that was both innovative by the method and by the result.

The Arrovian Social Welfare Functions are functions that takes as input all the individual preference orderings\footnote{A ranking is an ordering if (1) any two alternatives can be ranked and (2) the ranking to be transitive
\citep{sen2018collective}} over a set of \emph{Social states}\footnote{\emph{A
social state describe what is happening to the individuals and the society in
the respective states of affairs} \citep{maskin2014arrow}}, and returns a
\emph{social preference ordering}, which is the aggregation of the preferences of the individuals. The social preference ordering, as preference orderings, needs to be a complete preorder. To study the different possible voting schemes, Arrow proceeded to an axiomatic analysis. He defined 4 very reasonable axioms that every Arrovian SWF should respect \citep{maskin2014arrow}:

\begin{itemize}
    \item Unrestricted Domain (U): As the aggregation procedure is designed
        before individual preferences are known, the Arrowian SWF must work
        for any individual preferences. In other words, there are no
        restrictions in the ordering of the individual.
    \item The Pareto Principle (P): We need a social ordering that is
        consistent with the Pareto principle. In other words, if everyone
        strictly prefers $x$ to $y$, then $x$ must be preferred to $y$ in the social preference ordering.
    \item Independence of Irrelevant Alternatives (I): The social ranking of two alternatives $x$ and $y$ must only depend upon individuals
        preferences over $x$ and $y$ (regardless to any third alternative
        $z$). In other words, the way an individual ranks other alternatives than $x$ and $y$ is irrelevant as far as the social ranking between $x$ and $y$ is concerned.
    \item Non-Dictatorship (D): There is no dictator \emph{i.e.} there is no individual such that the social ranking coincides with this person's strict preferences, independently from the rest of the society.
\end{itemize}

Arrow's Theorem shows that if we have three or more alternatives and two or
more individuals, we cannot find a voting scheme that satisfies (U), (P), (I)
and (D) and which is an Arrowian SWF at the same time.

This result may be seen as surprising, but \cite{vincke1982arrow} shows that in fact it is not. The reason is that any SWF is a preference aggregation
procedure and as such will provide a result which is poorer (from an
information content point of view) from the aggregated preferences. In other
terms if we aggregate complete orders (such as weak orders) any SWF satisfying Arrow's conditions cannot yield a result with the same type of information content (a complete order). Not surprisingly Condorcet's procedure (which satisfies Arrow's conditions) does not guarantee the existence of a social complete order.

For \cite{sen1977social}, Arrow's impossibility result can be interpreted as a proof of Arrowian SWF informational limitations. Indeed, only considering
preference ordering and imposing strict no-interpersonal comparison is not
sufficient to be able to take decisions over social states.
For this reason, some welfare economists have developed SWF allowing interpersonal comparisons (for instance
\cite{adler2012well}, \cite{adler2019measuring}).

A criticism made by \cite{sen2018collective} is that Welfarists only consider
preferences (or utilities) of individuals independently from the nature of the social state that is considered. This is called the ``neutrality'' and it is a result of the axioms (U),(P) and (I). It implies that if a person is decisive on the determination of a preference between two social states, then this person should be decisive in every situation. In other words, the social order depends on individual preference only, and not on the \emph{nature} of the alternative. Neutrality is a lack of information and can lead to moral issues. To illustrate this problem, \cite{sen2018collective} gives the example of a cake division problem with two individuals preferring the largest share possible. It can be argued that if the cake is shared in the form $a$: (person 1: 99\%; person 2: 1\%), then the preference of person 2 should prevail on person 1 (because of an ethical argument). But if we consider that person 2 preference prevails on person 1 in $a$, then, due to ``neutrality'', person 2 prevails on person 1 regardless of the social state. So situation $b$ (person 1\%: 1; person 2: 99\%) should be socially preferred to (person 1: 50\%; person 2: 50\%). This is of course not ethical, and the way we aggregate individuals preferences should depend on the current social state. In order to consider the current social state, we need more (not necessarily utility) information \citep{maskin2014arrow}.

%\subsection{Capability} THIS IS REDUNDANT. WE ONLY NEED ONE SECTION ABOUT CAPABILITY.
%
%\subsubsection{Why CA}
%
%The CA has been greatly influenced by the welfare economics,
%but also by John Rawls' conception of justice that is introduced below. The
%capability theory, has been developed from both approaches, and their
%criticisms.

\subsubsection{A Theory of Justice}

In \emph{A Theory of Justice}, John \cite{rawls2009theory} was concerned with
the problem of defining a fair society under the difficult question of
distributive justice. For him, the concept of justice is linked, and must be
achieved by fairness, which can be seen as a demand for impartiality. He wanted to define some basic structures, that would allow to have the greatest freedom and equality. Using an abstract reasoning, he defined the best basic structures (which are the political, economical and social institutions), to distribute rights and advantages that are the result of the social cooperation. To do so, Rawls got his inspiration from the social contract theory that emerge in the Enlightenment. He concluded that a fair society should maximize \emph{primary goods} of the least well-off. The idea of primary goods is a very important concept for Rawls. It is a list of what all citizens desire, no matter what else they desire. In \cite{rawls1982social}, he clarifies the notion of \emph{primary goods} and characterises them into five groups:

\begin{itemize}
    \item (1) The basic liberties, that can be defined as a list, such as the
        freedom of vote, the freedom of thought and speech...
    \item (2) The freedom of movement and choice of occupation.
    \item (3) The powers of offices and positions of responsibility, and in
        particular those associated with the political and economic
        institutions.
    \item (4) The income and wealth.
    \item (5) The social base of self respect.
\end{itemize}

The distributional concern is then on those primary goods, not on utility. This avoids problems of interpersonal comparisons that are faced by the welfarist tradition. To reach his \emph{maxmin} strategy, Rawls followed this simplified reasoning.

First, he defined an \emph{original position}, in which rational and free
individuals will have to design the institutions at unanimity in order to make a fair society. These people are assumed to be under a \emph{veil of
ignorance}, which means that they have no idea of the position they are going
to occupy in the society. It is a ``pure'' state of equality and impartiality; they have no idea if they are going to be rich or poor, or what political opinion they will have, and what is going to be their gender or mental and physical conditions.  \cite{rawls2009theory} argued that from the discussion of individuals, two principles will emerge:

\begin{adjustwidth}{30pt}{30pt}
\emph{First Principle: Each person is to have an equal right to the most
extensive total system of equal basic liberties compatible with a similar
system of liberty for all.}

\noindent \emph{Second Principle: Social and economic inequalities are to be
arranged so that they are both: (a) to the greatest benefit of the least
advantaged, consistent with the just savings principle, and (b) attached to
offices and positions open to all under conditions of fair equality of
opportunity.}

\end{adjustwidth}

The second principle can be divided into two points.

First, what is called the \emph{Difference Principle} (a). He claims that,
under the veil of ignorance, people focuses on the distribution of primary
goods (group (3), (4) and (5)), that are the basis for interpersonal
comparisons. Unlike the basic freedom, Rawls allows some inequalities on the
social and economic ground.   Under the veil of ignorance people seem to be
egalitarian. The only way inequality would be accepted is if it advantages the worst-off, because of the unanimity rule.

The second point of the second principle (b) emphasises the fact that
individuals having the same talents and willingness should have the same
educational and economical or political opportunities, indifferently of their
gender and family they were born in (no discrimination over religion, ethnicity ect).

\cite{rawls1982social}, gave the priority to the principle 1 over the 2, and to part (b) over (a). The first principle emphasises the fact that all individuals must have the same primary goods (group (1) and group(2)). Rawls placed freedom at an absolute priority.

One important difference between the welfarism and Rawl's theory is that the
first focuses on \emph{ex post} outcomes (it evaluate the possible outcomes)
whereas the later focuses on \emph{ex ante} opportunities (it gives bases for a fair society).

\subsubsection{Criticism of welfarism and Rawls's theory}\label{critic}
The CA emerges from a new interpretation of egalitarianism in Sen's \emph{Tanner Lectures: Equality of what} \citep{sen1979equality}. This
lecture starts with a criticism of welfarism and Rawl's theory.

\emph{1. General critique}. Utility is generally defined as happiness or desire fulfilment, and can be seen as a good representation of the welfare of an individual. Indeed, a person that is not happy, or that is not able to fulfill his or her desires will have a low level of welfare. Using only utility as a base for evaluation of socials state can be perceived as reasonable. Nevertheless, very strong criticisms can be moved to this approach which we can summarise in 4 points.

\begin{enumerate}

\item \emph{Subjective differences.} A first critisism made in \cite{sen1985commodities,sen1999development} is that utilities are subject to mental adjustments. A person (1) living in poor conditions can have ``realistic'' expectations and be happy havinge few desires to be fulfilled. Now, consider a person (2) from the upper class that would have very high expectations or desires, and claims to be unhappy, or doesn't have his/her desires fulfilled. Then, in the context of interpersonal comparisons (utilitarian), the utility of (1) is considered as greater than the one of (2). But someone observing how (1) and (2) are living would hardly claim that (1) well-being is higher than (2).

    On the other hand, by considering no interpersonal comparisons at all, the ``new'' welfarism using individual preference orderings, gives us no direct information about the welfare of the individuals. For example, given two individuals with the same preferences over some bundle of goods may not equally derive welfare from a bundle, due to the heterogeneity of individuals or environmental aspects. There is no reason that the same bundle would procure them the same level of welfare.

\item \emph{Distribution effects}. A second criticism made to the welfarism and CBA is the distributional indifference. This criticism is shared by Rawls and is linked to the notion of justice and fairness. The classical utilitarian, using the sum of utilities, would typically disadvantage people deriving less from resources than others. Let's imagine a person (1) with an handicap that makes him/her deriving less utility from resources than a person (2). As classical utilitarians are only concerned with the sum of utilities, they will give more commodities to person (2), who has a high deriving rate \citep{sen1979equality}. This may not be our moral intuition as the person with a handicap will be disadvantaged two times: first from his/her handicap and, secondly from the resource allocation. In the CBA, the people derive utility from income at the same rate, which, for the reason above seems unrealistic.

%On the other hand we have seen that ``new'' welfare economics are not able to
%discriminate situations that are Pareto efficient. The Nero example, where
%one's utility increases at the cost of thousands' decreasing, is not seen as
%worst than the initial situation. It is a clear distributional indifference.
%
\item \emph{Society Motivations.} The attempt of the Social Welfare Functional approach to introduce some moral aspects to deal with distributional issues, is not sufficient if we only consider utilities. Some very important non-utility based aspects of human life can be in contradiction with only utility based framework. For example, \cite{sen1970impossibility} proves that the Pareto principle is in conflict with the respect of personal liberties (that \emph{is being 'decisive' over some personal matter}), both being desirable. To illustrate this idea, we can give the following example inspired from  \cite{sen1970impossibility} and \cite{sen2009idea}. We have two individuals, (1): the prude and (2): the impish. They are proposed to watch a horror movie, and we can distinguish three alternatives; \\
    $a$: the prude (person 1) watches the movie. \\
    $b$: the impish (person 2) watches the movie. \\
    $c$: non of them watch it. \\
    The ranking of person 1 is $c,a,b$ (prefering to watch the movie rather than knowing that person 2 is enjoying himself in front of it). The ranking of person 2 is $a,b,c$ (prefering that person 1 watch the movie because s.he knows that it will chocked him/her, which is funnier than just watching the movie alone). From the liberal perspective, we have $c$ preferred to $a$ (because person $a$ doesn't want to see the movie), $b$ preferred to $c$ (because person 2 wants to watch the movie). So, from the liberal perspective, we have the order $b, c, a$. But this ranking is not coherent with the Pareto criteria (as they both prefer $a$ to $b$). This shows that if we only use utility we can get solutions that do not respect some values of a society, such as the freedom of choice on personal matters.

\item \emph{Individual motivations}. Finally, utility and CBA do not take into account that individuals may act for different purposes than their own interest. As argued in \cite{rawls2009theory}, individuals ``have a sense for justice'' and ``a conception of good'', that can lead them to act against their own utility. For example, one can start a hunger strike at the cost of being unhappy or desire unfulfilled for what s.he considers as ``the greater good''. Sen claims that using utility only, we cannot see the difference between someone that is fasting (for political or religious reason) and someone that is starving (without choice), whereas their situations and welfare are very different. An individual cannot, and should not be reduced to an agent that would only maximises its own utility, but rather as a complex individual which decisions are influenced by many factors.

\end{enumerate}

\emph{2. Critique to Rawls}. Concerning Rawls' approach, the first criticism made in \cite{sen1979equality} is that Rawls focuses only on primary goods. He argues that they are means and not ends (that are freedom), and that what we have to look at are the ends. Moreover, Sen argues that focusing on primary goods in a society where people are different can lead to unfair situations. Indeed, Rawls focuses on primary goods and not on what people can do with them. For example, a person that has a handicap may need more goods than someone else to achieve the same level of welfare, having the same primary goods don't guaranty having the same welfare.

Another criticism in \cite{sen2009idea}, is the fact that Rawls' theory is part of  the contractarian tradition \citep{hobbes1651leviathan,
rousseau1962contrat}. Sen views is that they are ``transcendental
institutionalism'' approaches. This term is composed in to part
\citep{thomas2013sen}; first, the \emph{transcendental} refers to the idea of
finding a set of perfect principles of justice, secondly,
\emph{institutionalism} underly the fact that the scope of distributive justice is limited to institutions only. He criticises the fact that Rawls is only concerned about \emph{just institutions} and not \emph{just societies}. The justice in a society can depend of non-institutional features, such as social interactions, or behaviours of people. Sen makes two principal criticisms on the transcendental institutionalism.

\begin{enumerate}
\item \emph{Feasibility.} First the \emph{feasibility} problem: \cite{sen2009idea} contests the fact that Rawls' original position will lead to the two unique principles. For instance, in \cite{roemer1998theories}, it is shown, using von Neumann–Morgenstern utility functions, that a \emph{maxmin} strategy (the difference principle) will be chosen only if individuals under the ``veil of ignorance'' were infinitely risk-averse, which is a very strong assumption.

    A unique solution from the original position presumes that there is only one impartial argument. But why would we have only one impartial solution, and how to be sure it leads to the two principles? Or, can individuals in the original position agree at the unanimity, on one set of principle? For Sen, multiple impartial arguments can coexist, which would ``invalidate'' Rawls' theory. To illustrate this idea, Sen gives its famous example of the \emph{three children and a flute}, consisting into choosing the child to give a flute, based on an impartial reasoning and the speech of the children;

\begin{itemize}
    \item \emph{Child 1} argues that s.he is the only one that knows how to
        play flute.
    \item \emph{Child 2} argues that s.he is the poorest, and is the only
        that has no toy.
    \item \emph{Child 3} argues that s.he just finished to build the flute,
        and doesn't want to give it.
\end{itemize}

It is very difficult to say which of those three children merit the flute. We can make different impartial arguments that would be in favour of each of them. Some utilitarians will be in favor of \emph{child 1}; ``s.he will gets the higher utility because s.he knows how to play''. Some other utilitarians will be in favour of \emph{child 2}; ``s.he is so deprived that even if s.he doesn't know how to play, having the flute will increase a lot his/her utility''. Egalitarians will also be in favour of \emph{child 2}. Finally, libertarians will be in favours of \emph{child 3}; ``workers should be able to enjoy the fruits of their own labour''. It seems complicated to define a unique impartial argument. Then, how individuals could agree on a unique transcendental agreement?

Moreover, even if such agreement is found, it is not guarantied that citizens' political conception of justice will automatically accommodate to the new social state. Indeed, from a practical point of view, we cannot guarantee that all the individuals will agree on the chosen political conception.

Finally, defining principles, institutions and rules is different from the real social state in the society produced. We have to evaluate if the produced society is actually fair. If it's not, we are in a deadlock because we already have defined perfect institutions...

\item \emph{Redundancy}. The second criticism to the contractarian tradition is \emph{redundancy}: the non possible realisation of the perfect identified social arrangement. The issue is that, if the perfect solution is unreachable, the transcendental institutionalism approach is useless, because it will not help to evaluate possible policies. Indeed, knowing that someone's favorite food is sushi, will not help us to know which plate s.he is going to choose at an Italian restaurant. One could argue that if we know the ideal solution, we only have to choose the ``closest'' possible solution. But, first, it's difficult to calculate the distance between alternatives that can be different on multiple dimensions. And secondly, the ``closeness'' is not necessarily linked to its appreciation. I may prefer yellow paint than cyan, and still prefer cyan paint than green (even if green is closer to yellow as it is a mix of yellow and cyan paint).

\end{enumerate}

\subsection{What is the capability approach?}

\subsubsection{Equality of what?}

In \cite{sen1979equality} and later in \cite{sen2009idea}, it is argued that
the conception of justice is based upon the one of equality. We therefore,
need to answer the question ``equality of what?'' For instance, utilitarians
focus on the equality of the marginal utility. Indeed, maximising the total sum of utilities is the same as equalising individual marginal utilities (as
utilities are considered as concave).

For Rawls, what has to be equalised are primary goods, or equality of
opportunities. We should accept inequality only if it profits to everyone, and if it respects the basic liberties, freedom of movement and chance of
occupation.

Others have focus on equal revenue, equal capital, equal redistribution of the value created  etc. For Sen, what we should promote is equality of the
``capabilities'', or the substantive opportunity and freedom.

\subsubsection{Capability Approach}\label{CA}

The capability approach has risen a lot of interest these last decades and a lot of different researchers have worked on this subject. The result is the existence of several, slightly different, conceptions of capability and different terminologies. Even Sen's own writing has conceptually and terminologically changed over time. In this paper we will mainly focus on his presentation in \cite{sen1985commodities}.

According to \cite{sen1997choice} and \cite{sen1985commodities}, we should do a distinction between owing a good, using it and obtaining utility from its use. Welfarists, mainly focuses on income and/or commodities, but for Sen, while goods and income are important, we also need to know what we can do with them. First, a commodity has to be distinguished from its characteristics. For instance, having a laptop can give access to knowledge, social interaction, work, videos etc... But the characteristics of a commodity doesn't tell us what the individual will effectively do with it, or is able to do with it. For instance, we can do several different things with a laptop, but we usually only use a few of them. Moreover, some persons are not able to use all the properties of a commodity, or use them less efficiently. Individuals are different and can use the characteristics of a commodity differently. Such differences can be explained by different \emph{conversion factors} that an individual possess. According to Ingrid \cite{robeyns2005capability} these factors can be divided into three categories:

\begin{itemize}
    \item \emph{Personal conversion factors}: they are related to physical
        and mental conditions, sex and gender, reading skills etc. In the
        laptop example, the reading skills have an influence on the possible utilisation of a laptop
    \item \emph{Social conversion factors}: they are related to public
        policies and laws, the social group we are in, social norms,
        discrimination practices, power relations etc. For instance, the
        government can limit the access to internet for some people.
    \item \emph{Environmental conversion factors}: they are related to
        climate, geographical location, pollution etc. In our case, living in a region that is poorly or not connected to internet will limit the utilisation of the laptop.
\end{itemize}

Since looking at an individuals' commodities in order to establish her welfare is not sufficient, we have to look at their \emph{functionings}, which is \emph{what a person is actually able to do}, given his/her commodities. Watching a video or working on a laptop are two different functionings, that are different from owing a laptop.

%Functioning are generally seen as dimension of liberty, such as being healthy,
%having a good job, or more complex liberties such as being respected and being
%happy.

The person's \emph{capability} is all the possible combinations of functionings that an individual can reach. It has to be distinguished from the \emph{achieved functioning}, that is the set of ``doing'' effectively chosen by the individual. A capability is the ability to achieve, whereas a functioning is an achievement. Capabilities represent the effective freedom of an individual to choose between different functioning combinations, that s.he has reason to value. From a social choice theory approach, a capability can be seen as an opportunity set. Then, only the ends (possible combinations of functionings) have an intrinsic value in order to evaluate welfare. The means are only seen as instruments to reach those ends.

For a given combination of functionings, individuals have a ``being'' state,
which describes how somebody is (mobile, happy etc.) As the ``beings'' of individuals derive from their doings, functionings are usually seen as the ``beings'' and ``doings'' of an individual. They are on their own morally neutral, they just describe how an individual act and is.

We have to make the distinction between the realisation of a (or a combination) of functioning(s) and their valuations. Two persons can watch the same video on the same laptop, but they can have a very different appreciation of it. Those valuations are influenced by individuals' \emph{conversion factors}, but also by their personal history, values, desires etc. Indeed, people with the same capability set, will not necessarily choose to achieve the same functionings because of their different ``ideas of the good life''.

Moreover, individuals are considered as \emph{agents}, which means that they
can act in order to change, regarding their values and goals. We can imagine
that they will not necessarily choose to achieve a functioning that maximises
their happiness because they respond to complex reasonings that are represented in their \emph{valuation vectors}, and they may choose to act to change the society at the cost of their happiness. To illustrate this idea Sen uses the example of an individual who is fasting for political reasons at the cost of his or her health and happiness.

One of the first efforts of Sen has been to develop his own framework to
represent capability  \citep{sen1985commodities}:

\begin{itemize}
 \item[$X_i$:] Individual's entitlements as described in \cite{sen1982poverty}, is the set of bundle of commodities that can be
     chosen by the individual through legal channel. It depends on initial
     individual's ownership and the set of bundles that s.he can obtain from
     trade or production.
 \item[$x_i$:] The vector of commodities possessed by individual $i$
     \emph{i.e. 5000\$, a car, some food...}. This vector is chosen in the set of his/her entitlement.
 \item[$c(x_i)$:] being exogenous function converting a commodity vector into a vector of characteristics; $c(x_i)$  is the characteristics
     of the consumption of $x_i$ \emph{i.e. spending power, possibility of
     displacement, nutritional aspect ...}.
 \item[$f_i(.)$:] Person $i$’s “personal utilization function”. It is a
     function that converts certain characteristics into functionings using the \emph{conversion factors}.
 \item[$F_i$:] The set of all ``utilisation functions'' $f_i$ that $i$ can
     choose in.
 \item[$b_i = f_i(c(x_i))$:] The 'beings' of a person that s.he accomplished using the commodities $x_i$ and a utilization function $f_i \in F_i$.
 \item[$v_i(b_i)$:] being the valuation function. $v_i(b_i)$ is the
     valuation of the beings $b_i$.
\end{itemize}

%\begin{tabularx}{\textwidth}{llX}
%     $X_i$ & $:$ & Individual's entitlements as described in
%     \cite{sen1982poverty}, is the set of bundle of commodities that can be
%     chosen by the individual through legal channel. It depends on initial
%     individual's ownership and the set of bundles that s.he can obtain from
%     trade or production. \\
%     $x_i$ & $:$ & The vector of commodities possessed by individual $i$
%     \emph{i.e. 5000\$, a car, some food...}. This vector is chosen in the set
%     of his/her entitlement.\\
%     $c(x_i)$ & $:$ & being exogenous function converting a commodity
%     vector into a vector of characteristics; $c(x_i)$  is the characteristics
%     of the consumption of $x_i$ \emph{i.e. spending power, possibility of
%     displacement, nutritional aspect...}\\
%     $f_i(.)$ & $:$ & Person $i$’s “personal utilization function”. It is a
%     function that convert certain characteristics into functionings using the
%     \emph{conversion  factors}.\\
%     $F_i$ & $:$ & The set of all ``utilisation functions'' $f_i$ that $i$ can
%     choose in. \\
%     $b_i = f_i(c(x_i))$ & $:$ & The 'beings' of a person that s.he accomplished
%     using the commodities $x_i$ and a utilization function $f_i \in F_i$ \\
%     $v_i(b_i)$ & $:$ & being the valuation function. $v_i(b_i)$ is the
%     valuation of the beings $b_i$.\\
%\end{tabularx}\\

For individual $i$, given a vector of commodity $x_i$, the set of
\emph{possible} functionings is: $$P_i(x_i) = \{b_i | b_i =  f_i (c(x_i)),
\text{for some} f_i \in F_i\}$$

Then the \emph{capability} is given by:
$$Q_i = \{b_i | b_i = f_i (c(x_i)), \text{for some } f_i \in F_i \text{ and some} x_i \in X_i\}$$

This mathematical representation of capability is rather a formalisation to
understand the relation between the different elements of the theory
(\emph{i.e. what is the link between commodities, characteristics, individual
conversion factors, functionings, capabilities...}) than a practical formula
that could be used to represent individual's capabilities.

One of the strengths of the capability approach is that it accounts for human
diversity. On the one hand, it allows to consider the variation of conversion
factors of a commodity characteristics into functionings. Some of those
conversion differences are due to individual factor differences and other are
more structural \citep{robeyns2003capability}. For instance, one is able to
eat pasta, where another cannot, due to gluten intolerance, making the
conversion factor of pasta into lower pleasure or lower wealth. One more
structural variation into conversion factors, is the difference between men and women as far as the conversion of work into income is concerned. This second conversion difference can be used in order to reveal social inequality.

On the other hand, the capability approach allows to take into account
multidimensional diversity of humans life. Inequality or deprivation can occur on very different dimensions. For example, restricting sex inequalities into income study is not sufficient. We have to look at functionings such as ``being free from sexual harassment'' or ``being free from household work''.

\subsubsection{Difficulties and criticism of the capability approach}
%NOT COMPLETLY SOLVING INTERPERSONAL

Let's begin with a remark that has been made in \cite{robeyns2000unworkable}:
Sen's CA is not completely solving the \emph{interpersonal comparison} problem. Most of the capabilities  allow  interpersonal comparisons, especially the basic ones. Having access to safe water, shelter, and being sufficiently nourished can be considered ``objective capabilities'' since there is a social consensus about their importance for any citizen's wellbeing in any part of the world. We do not really need to assess their subjective value and they can be objectively observed. Nevertheless, if we consider some more complex functionings such as access to culture, urban quality, self-respect (just to mention some very diverse), it seems much more difficult to do ``objective'' comparisons. Capabilities containing subjective judgements cannot be fully compared between different individuals. Indeed, such ``subjective functionings'' are not only influenced by the personal, social and environmental factors (in which individuals have not complete control), but also by their values. In other terms in order to compare such capabilities we need to know the subjective value citizens give to certain achievements and/or opportunities: we need to enhance the information basis of commensurability. In any case, there is still room for some limited interpersonal comparisons, of complex capabilities, and more generally for capability set comparisons.

One of the most widespread comments is that CA is too \emph{individualistic} \citep{deneulin2010capability,stewart2002amartya, stewart2005groups,kaushik2011functionings}, mainly because individuals are atomized. It is clear that considering individual's capabilities, the utilisation of Commons, social structures and the interactions between individual raises important issues as far as the real \emph{individual} freedom measurement is concerned. Indeed, for Commons, the substractable effect implies that the ``real'' capability of an individual will depend on the utilisation of the Commons by other individuals. But there is not intrinsic impossibility in the capability approach to take into account those social environments. Most of the work referring to the CA has not paid a lot of attention to groups, but some did: \cite{kynch1983indian}. Besides, social aspects are theoretically taken into account as social conversion factor of resources into functioning. See \cite[Chapter~4]{robeyns2017wellbeing} for more discussion on the individualistic argument.

Another criticism that have been made is that the CA is not operational.
This is due to different difficulties, from the lack of a list of capabilities, the lack of frameworks to compare different capability sets, the access to data and an utilisable framework to represent capabilities. The CA is neither an income-based, nor an outcome-based theory, but rather an opportunity-based approach. In the two first approaches, observations are possible because we are concerned by the actual individual lifestyle. One of the major issues for an empirical use of the capability approach is that it has to measure the potential lifestyles (which by definition, don't exist).

\cite{sen2009idea} argues that public policies should focus on equalising the
individuals' capabilities, but he did not provide a \emph{framework to compare} them. How to compare someone that has the opportunity to be in good health but is not wealthy with someone that has the opportunity to be wealthy, but is not in good health? As the capability approach leads to multi-dimensional comparisons, it can be hard to find ways to compare different capabilities. Considering that functionings are fully comparable, an easy way to compare two capabilities would be to have:

$$Q_i\succ Q_j \text{ if } \forall
b_j \in Q_j \implies \exists b_i \in Q_i \text{ such that } b_i \text{ Pareto
dominates } b_j .$$

However, the Pareto criteria is not the only possible rule to determine binary relations.  Some other rules, being less strict in terms of dominance, could be introduced and may be developed in future works. Moreover, Sen claims that we should not equalise capabilities at every cost. For example, if women live longer than men all other things being equal, then we should not reduce women's access to hospital to make men's and women's life longevity equal. Then, both the difficulty to compare capability and the
unanswered question of ``what exactly do we have to equalise?'', raise issues
on the operational use of the capability approach.

The \emph{access to data} is another limit to the concrete utilisation of
capabilities. Indeed, in order to use the CA operationally, we need a lot of different data. First we have to find, for each individual, their means in order to achieve their functionings; their income, commodities, access to Commons etc, as well as the market mechanism. Data for this first stage are more difficult to find that in some ``income approach'', which deals with net income only. Then we have to collect information about the conversion factors of individuals (\emph{i.e. personal, social and environmental factors}). And finally, we have to find their ``values''. Most of data that have been used so far are second hand data that have not been specifically designed by CA scholars. Besides, there is no formalisation that is both operational and that fully represents the capability approach. In fact, as argued by Robeyns, the informational richness that is needed, is not only a data collecting problem, but also a mathematical modelling one. Social constraints and different values are, among other things, difficult to model. While using a mathematical representation, we have to be careful to not lose the richness of the capability approach by oversimplifying.

Some have criticised the \emph{Under-Theorisation} of the capability approach. The fact that Sen doesn't want to give a list  of functionings (unlike \cite{nussbaum2003capabilities}), makes the theory unclear and hard to use. This come from the fact that the capability approach is not normative. From the \emph{context of decision aiding process} \citep{tsoukias2007concept}, the method is the key element and the selection of particular functioning can be leaved to a discussion with different stockholders. Not having a list of definite functioning allow a framework to be \emph{versatile}, which is desirable from the point of view of decision makers.

%\begin{note}
%Is the CA too individualistic?\\
%Is CA reinventing the wheel ? Social science uses multiple data before Sen...\\
%Data set used in empirical word are generally second hand data that have not
%been specifically design for CA.
%\end{note}

%We think that a mathematical formalisation in view of operational application
%is still worth it. Indeed, as we will see in the following section, capability
%theory is applied in the real world and we have experience in using it for
%policy recommandation purposes. Under such a perspective we lack a theoretical
%framework within which compare, analyse, discuss such experiences and then
%further build the theory.

\begin{landscape}% Landscape page
\begin{table}[]
    \centering
    \begin{tabularx}{\linewidth}{l|XXXXXX}
         & Index & Fuzzy & Statistical & Structural equation & Clustering & Conceptual framework \\\hline
         Inequality/poverty & \cite{UNDp2019,alkire2014multidimensional} & \cite{martinetti1994new,martinetti2006capability,cerioli1990fuzzy,cheli1995totally,qizilbash2002note,qizilbash2005capability,martinetti2000multidimensional}& \cite{lelli2001factor,schokkaert1990sen,ram1982composite,klasen2000measuring,morris1979measuring}& \cite{addabbo2004extent,di2007children}& \cite{zeumo2014new} & \cite{sen1985commodities} \\\hline
         Health& \cite{coast2008valuing,flynn2011assessing,al2011estimation} & & & & \\\hline
         Urban & \cite{blevcic2013capability,blevcic2015evaluating, blecic2015walkability}& & &&\cite{Fancellolurning,FancelloetalSEPS2020} &\cite{frediani2008planning, frediani2012processes,nordbakke2013capabilities} \\
    \end{tabularx}
    \caption{Application of CA by subject and method.}
    \label{tab:referecne}
\end{table}
\end{landscape}

\subsubsection{Applications of the capability approach}\label{app}

As argued in \cite{robeyns2003capability}, we can distinguish three levels in which CA can be used;

\begin{enumerate}
\item \emph{A critique}: As shown previously, the capability approach comes
    from a critique of former ways of evaluating well-being, such as wealth
    or welfarism, as well as a critique of Rawls’ theory of justice.
\item \emph{A paradigm}: It can be seen as a framework of thought giving
     information about an individual’s advantage and the social arrangement of a society. It can be used to measure poverty or development, which are fields particularly investigated by Sen, as well as an informational tool to promote new public policies. The focus on different informational bases, rather than only using wealth to evaluate poverty can be seen as a consequence of a change of paradigm raised by the CA.
\item \emph{A Tool to make interpersonal comparison of well-beings}: A procedure (a formula or an algorithm) that allows us to make interpersonal comparisons.
\end{enumerate}

These levels are interconnected, the paradigms coming from the critiques of
Rawls' theory, and formulas and algorithms can be used to study empirically
poverty and inequalities. As the first level has already been discussed, in
this section we will see some applications of the capability approach in the real world (using both level 2 and 3).

Table \ref{tab:referecne} is an overview of the literature on the operational aspect of the CA. We identify three major subject of studies; inequalities and poverty identification, health and urban planing. Different methods have been used to measure capabilities;

The first two empirical use of CA have been made by Sen himself in
\cite{sen1985commodities}. He shows that the GDP is not a sufficient tool in
order to study inequalities. His first ''case studies'' focus on the
''development level'' of different countries. Comparing Brazil and Mexico to
China and Sri Lanka, Sen shows that even if the GDP per capita of the first is about 7 times higher than the second, life expectancy is higher and infant
mortality is lower in Sri Lanka and China. This CA used different measures of human functionings, in educational, health and economics fields, showing that the order obtained considering the GDP is very different from the one obtained, for instance, considering life expectancy. He shows that the GDP is an imperfect indicator of human development. In his second application, Sen studied sex inequalities in India, showing that women have ”lower health level” than men, suffering more from morbidity or malnutrition.

The Human Development Index (HDI) is generally cited as an example of \emph{index}  that derives from capability approach. This index measures human development through three different indexes that can be seen as fundamental capabilities: the Life Expectancy Index (LEI), the Educational Attainment Index (EAI) and the Adjusted real GDP per capita Index (PPP). The HDI score is obtained through a two-steps procedure: first, the parameters on the different dimensions are re-scaled using the minimal and maximal values in order to obtain values between $0$ and $1$. Then the HDI is obtained using the formula $(LEI\cdot EAI\cdot PPP)^{1/3}$ \citep{UNDp2019}. It is one of the first large utilisations of a composite index, that focuses on different dimensions of life, and it shows once again that the GDP should not be the only indicator of human development.

Another well-known composite index is the Multidimensional Poverty Index
\citep{alkire2014multidimensional}. One of its specificity is that it uses
small data-sets (household surveys). It is composed of 3 dimensions: Health,
education and standard of living. Each dimension is an aggregation of yes/no
questions (giving 1/0 score). Then, a weighted sum is applied and a household
is considered as poor if it is under a certain score (the poverty line).

Non statistical techniques have also been used in order to measure urban
capabilities \citep{blevcic2013capability}. In \cite{blecic2015walkability}, a score is given to all points of a city (Alghero, Italia). The score describe its ``walkability'', or, jointly how many destination (commerce, service, leisure) are reachable from the point, at what distance and what is the quality of the pedestrian path. Different scores have been calculated regarding different profiles (having different ``utilisations'' of the city), for instance tourists or parents. In \cite{blevcic2015evaluating}, the same
framework is used in order to find good potential pedestrian routes in Lisbon
(Portugal).

Other applications of the CA can be found in health economics. Several measures have been developed such as the ICECAP-A \citep{coast2008valuing}, a measure of capability for the adult population, the ICECAP-O \citep{flynn2011assessing}, a measure of capability for older people,
or the CES \citep{al2011estimation}, a Carer Experience Scale measure of
care-related wellbeing. Those indexes are based on the same procedure. First a questionnaire is constructed with the population to be observed, thanks to a qualitative work. Then values of the population are learned through
quantitative work. Finally, individuals respond to the questionnaire and a
score is calculated with weights learned in the previous step.

%klasen2000.pdf(Two procedures were used to derive a weighting of the various
%components of the index. One derived the weights from the data itself based on
%principal component analysis.13The other is to calculate the total deprivation
%index as simply the average score of all individual components. It turns out
%that the two procedures yield virtually identical results. The correlation
%coefficient between the deprivation measure arrived at by the two procedures is
%0.992.)

The \emph{fuzzy} set approach is an extension of the scaling method. Scores obtained
in different dimensions (or sub-dimensions) depend of a poverty membership
functions \citep{martinetti1994new, martinetti2006capability}. For instance,
considering a functioning such as being nourished, an individual (or household) will be given 0 if s.he is considered as really deprived (\emph{i.e.} starving), 1 if s.he is not deprived (\emph{i.e.} well nourished) and a value between 0 and 1 if s.he have some nutritional deficiency. The membership function can be determined by common sense and experience \citep{cerioli1990fuzzy}, or using linear or nonlinear function (such as sigmoid or logistic curves). Another technique is to use Totally Fuzzy and Relative approach \citep{cheli1995totally}, that gives a membership function  fully depending of the individual's relative position in the distribution. Concerning the aggregation of different dimensions, they use variants of the fuzzy union and intersection.  See \cite{qizilbash2002note} and \cite{qizilbash2005capability} for a fuzzy set theory application on south African countries and \cite{martinetti2000multidimensional} in the Italian context.

Non-statistical methods have the advantage to be easy to interpret and
understand. They are based on the maximum achievable functioning level and a weighted aggregation of measures. The drawbacks are that the maximum achievable functioning level and weights can be hard to find \citep{kuklys2004measuring}.

Then we have several statistical method such as the Factor Analysis and the Principal Component. Let's start with the \emph{Factor Analysis}. This approach is based on co-variance analysis. Its main goal is to summarize and reduce data
by presuming correspondence between different functionings \citep{lelli2001factor}. In \cite{schokkaert1990sen}, a study is led on around 500 Belgian Unemployed citizen. A factor analysis has been done on
questionnaires containing $42$ functionings. They reduced the number of
functionings to $6$ basic functionings that negatively influences individuals
welfare : social isolation, general feeling of happiness, bad physical
functioning, microsocial contact,  degree of activity and financial problems.
They showed that income loss, age, sex and family composition matter much more on their basic functionings than net disposable household  income  per month.

The \emph{Principal Component Analysis} is also based on the co-variance. It consists on computing linear combinations of original functionings, capturing
successively the proportion of the original functionings variance
\citep{ram1982composite}.  The principal component is the linear combination
"explaining" the largest part of the variance. The second component explains
the largest part of the remaining variance ect. In \cite{ram1982composite},
Principale Component Analysis computes new weights on the \emph{physical
quality of life index} \citep{morris1979measuring}, that is calculated with
life expediency index, infant mortality and adult literacy. The new weight
obtained from the (first) principal component captured about $95\%$ of the
variance. See \cite{klasen2000measuring} for another application on the south
African region. One of the advantage of this method is that ``weights'' are
found from the data. Statistical methods have the advantage to help to show causal aspects; for instance by underlying most important factors of deprivation. The structural equation method \citep{addabbo2004extent,di2007children}, is another method to find causes of inequality.

Clustering can also be used in CA framework. The meaningful
multidimensional poverty measurement \citep{zeumo2014new} follow a decision
aiding perspective, and has been used on 1255 households of Ouagadougou
(Burkina Faso). It is a two stage procedure. The first stage consists to an
unsupervised clustering (based on minimization of variance) of a population to form homogeneous socio-economic commodities class. Eight cluster have been
found using 48 socio-economic variables. Then capability set are assigned to
each cluster. Capability being composed of different functioning vectors,
themselves being level of achievement (good, average, bad) on 6 welfare
dimensions (nutrition, education, water and sanitation, housing, health and
transportation). The second stage consist to design policies to improve the
capability set of one or several cluster. In \cite{Fancellolurning} and \cite{FancelloetalSEPS2020}, a methodology is proposed to define clusters of citizens with similar preference and value over Commons, aiming to help to design public policies to improve urban capabilities.

%Finally, structural equation modelling have been used in the measurement of
%health and housing capability in the UK \citep{kuklys2004measuring}.

Despite the extension of applications we can observe some weaknesses. First, most of operational works have focused on present individual \emph{functionings} rather than \emph{capabilities}. The empirical work on CA being dominating by poverty measurement/identification, it is not a surprise. Functionings can be considered as more relevant than capabilities in this context; being undernourished is rarely a choice \citep{robeyns2006capability}. Choosing other applications and contexts, focusing on functionings can appear as less efficient. For instance, if someone wants to \emph{predict} the welfare of citizens of a ``developed country'' after implementing a public policy, measuring future \emph{functionings} imply assumptions on what are going to be the citizens' choices and their notion of a good life. In this context capabilities can make more sense because less prediction have to be done on citizen's future actions.

Citizens' welfare depends upon a complex bundle of goods and services, some (many) of then interacting between them and/or being part of more complex systems such as cities, communities, territories etc.. Designing a policy impacts the structure of such systems and therefore the welfare in ways which we do not always know. Under such a perspective the CA, as it stands, lacks a systemic vision: it has been mostly used to measure and identify or to find the causes of poverty/inequality and once again under a very simple deterministic representation of the phenomenon. Another consequence of the absence of a systemic vision is the difficulty to specify the impact of the Commons (and among these to identify the relevant ones) upon citizens' welfare. We will turn back to the benefit of a systemic vision of welfare in section \ref{CAisGood}.

We can note that low attention is usually given to individual subjectivity and values.
Ones again, this is probably due to the fact that most applications have been about poverty and inequality, as in this context all functionings are considered as crucially important \citep{sen1993capability}.
Nonetheless, the account for individuals values and perception is an important aspect of Sen's capability approach, and more attention should be given to it.
Moreover, the account for individuals' values is generally integrated in the weight for the aggregation of different functionings, especially in indexing method.
This can be misleading since the users and decision makers might not be aware of such hypothesis.
Taking into account the citizens' subjectivity and values should be integrated explicitly in the evaluation of functionings.

\subsection{Decision aiding and Capability Approach}\label{CAisGood}

Despite the CA is not really operational we consider it remains an interesting framework from a decision aiding perspective. We can identify four main steps in which a framework can be useful for decision aiding purposes, especially in the context of a public decision occurring in a policy cycle, within a policy analytics framework \citep{tsoukias2013policy,de2016evidence, daniell2016policy}.

\begin{enumerate}
    \item Help the client to have a better \emph{understanding} of the
    problem;
    \item Help to \emph{imagine and design} potential solutions;
    \item \emph{Explore} consequences of different solutions;
    \item Provide some \emph{arguments} in favour or against any selected solution.
\end{enumerate}

The reader should note that our aim is to use the CA as a common ground helping to rationalise how public policies are designed, assessed and implemented. Under such a perspective our vision differs from the mainstream proposal of the CA: some scholars (as \cite{nussbaum2011perfectionist}) consider only the governments to be the actors of policy design and/or improvement and think that the CA should address ``recommendations'' about public policies to such governments only. Yet, as stressed by \cite{stewart2005groups} and \cite[Chapter~4]{robeyns2017wellbeing}, improvement generally does not come
from the only benevolence of governments. It comes with political pressure from different groups. Clients such as NGOs, trade unions or economic actors can ask help for different purposes and aims, but they all need a common ground upon which discuss, negotiate and agree (if possible). In the following we will briefly discuss why the CA can be such common rationalising ground.

We consider three main characteristics for which the CA is useful for policy design purposes.

\begin{itemize}
    \item Introduces a \emph{multidimensional} approach of welfare;
    \item Accounts for citizens' \emph{subjectivity and value};
    \item Recognizes citizens' \emph{diversity}.
\end{itemize}

We also think that a fourth aspect should be developed in order for the CA to be fully efficient in a policy analytics context; a \emph{systemic} modelling. Table \ref{tab:sumupCAtoPA} summarise how the advantages of the CA can help the different steps of a policy cycle.

\begin{table}[h!]
    \centering
    \begin{tabular}{l|cccc}
    & Understand & Design & Explore & Argument \\\hline
         Multidimensional & \checkmark & \checkmark &&\checkmark\\
         Subjective & \checkmark & \checkmark \\
         Diversity  & & \checkmark & &
         \checkmark\\
         Systemic & \checkmark &\checkmark&\checkmark
    \end{tabular}
    \caption{How the CA is useful to construct a policy analytic framework.}
    \label{tab:sumupCAtoPA}
\end{table}

\paragraph{Understanding} A single figure representation of welfare is certainly easier to perceive and communicate. However, as already shown by many authors (for instance in \cite{sen2009idea}) such single figures conceal the rich picture of the citizens' welfare.

Welfare is a complex social, economic and cultural reality and a single figure will not be able to represent such a complexity. Moreover, most of the times the way through which such single figure is obtained conceals both arbitrary hypotheses contained in the aggregation procedure as well as an important differences among different dimensions of welfare (which could be compensated among them). A multidimensional representation of welfare offers a richer picture of the reality under observation, allows to see welfare as a distribution (and not as a figure) which on its turn allows to imagine alternative distributions in case this is considered necessary.

On the other side it is important to realise that welfare is perceived \emph{subjectively} and also assessed subjectively. Generally policies are expected to be a reply to a problem situation, but the extent to which the present distribution of welfare is a problem and for whom is a matter of subjective appreciation by each single citizen and the same idea applies as far as the impacts of any policy are concerned. Taking into account explicitly such subjective dimension allows to have a more realistic picture and to anticipate the different reactions of groups of citizens sharing common perceptions and values.

Last, but not least, adopting a \emph{systemic} approach as far as the representation of welfare is concerned, we allow taking into account the multiple interactions between access to private and public goods, access to the commons, private attitudes, individual and collective behaviours. Once again we obtain a richer picture upon which build a policy design.

\paragraph{Designing} Innovative policy design means: \\
 - being able to target specific categories of citizens in order to increase and improve policy legitimacy; \\
 - explore a space of solutions ``out-of-the-box'', avoiding dominant designs and creating new ideas and concepts; \\
 - anticipate the drawbacks and negative reactions improving efficiency and long-term acceptability. \\
A \emph{multidimensional} representation of welfare allows to expand the space of potential solutions including options apparently inconceivable, but potentially feasible. It also allows to explore deep ``what-if'' questions: what is needed in order to transform infeasible options to feasible ones or to make inconceivable actions realistic? Such an analysis is essentially possible only when the \emph{systemic} nature of welfare's definition and structure is explicitly considered as we suggest in Section \ref{model}. It is through the explicit representation of the multiple interactions between resources, actions and values that different designs become visible.

At the same time analysing the \emph{subjective} values driving the citizens' behaviour allows to identify different policy targets, to expand the inclusiveness of policies, while recognising the citizens' \emph{diversity} and their expectations helps in anticipating policy legitimacy and long-term acceptability.

\paragraph{Exploring} Rational policy design is possible only if we are able to offer a common ground where the consequences of different policies can be anticipated and measured. Policies should be simulated, projected upon the citizens and their impacts studied, including possible drawbacks and unforeseeable outcomes. A \emph{systemic} representation of welfare allows conducting such type of exercises and it is exactly for this purpose that we suggest a mathematical programming representation of the CA in Section \ref{model}. Such a model allows for quantitative analysis, to conduct simulations, to visualise impacts, offering to the stakeholders a common ground upon which discuss and negotiate (in case this results useful). A formal model is a necessary condition for any effective participative policy making process.

\paragraph{Arguing} Arguing for or against a policy (design, targets, objectives, consequences, measures, feedback etc.) is an essential feature for its legitimacy and effectiveness. Considering explicitly the \emph{multidimensional} nature of welfare as well as the \emph{diversity} of the citizens (with respect to the distribution and use of welfare) allows to construct the necessary basis for arguing effectively (this being a necessary condition for an effective participation to the policy design process). It also allows to identify those groups of citizens excluded by the policy designed. On the other hand it allows to construct recommendations based upon awareness, consciousness and argued convictions.

\section{Modelling capabilities}\label{model}

In order to show how the CA can be useful for decision aiding, we will present a framework aimed at ``measuring'' citizens' welfare. This framework could be applied in different public policies decision aiding context, but, without loss of generality, we will apply it to the context of \emph{modelling of collective threats}. We will see that it can be useful for a better understanding of the welfare's distribution, as well as in order to anticipate the impacts of different scenarios in case of Commons' damages and/or after introducing a policy.
%\begin{note}
 % If the topic of the paper is to show that capabilities are useful for
  %decision aiding then we need to change the above paragraph.
%\end{note}

\subsection{Introduction}
%\subsection{The problem}

We are interested into modelling collective threats. A \emph{threat} can be
defined as an event that has a certain likelihood to occur, and that will
produce a \emph{negative} impact on citizens' lives. More precisely, we
consider a threat which will have a negative impact upon a \emph{Common} and thus, affecting a population on a given territory.

A \emph{Common}, as defined by \cite{ostrom1990governing}, is a resource that can be ``natural'' or ``man-made'', and that is sufficiently large or/and too costly to be operated in a way that exclude potential beneficiaries from obtaining benefits from its use. Commons, that are also refereed as \emph{Common-pool resource (CPR)}, are said to be highly non-excludable, which means that a non-paying individual cannot be prevented from accessing it. As for public goods, they are shared by several people. The distinction made between those two types of goods is that the CPRs are highly
substractable: the utilisation of the resource by any citizen will prevent or degrade simultaneous utilisation of it by other citizens, or will consume part of the resource, reducing the quantity of this resource available for other individuals. ``Crowding'' and ``overuse'' effects are typical consequences of the substractable aspect of CPRs. Forests, fishing areas or roads are examples of CPRs.

We consider that citizens are negatively impacted in case their welfare is
``reduced''. The basic idea is that welfare is represented by the citizens'
capability sets and thus, a reduction should become visible through the
``reduction'' of such capability sets. A threat associated with a Common, such as the damage of a road or the pollution of a lake, is going to have a negative impact on citizens' lifes; for instance, by reducing their freedom of movement or reducing their leisure opportunity. These negative impacts are seen as capability deprivations. The negative impact of a Common's damage can be seen as the difference between the initial citizens' capability set, and its new deprived capability set.

Our approach consists in three distinct steps: \\
 1. at the first step we establish a generic model aiming at computing the capability set of a given citizen (a measure of his/her welfare as it stands presently); \\
 2. at the second step we use the same model in order to simulate the impact of an event altering the access to the Commons for a given citizen (and thus, altering the present distribution of welfare); \\
 3. at the third step we cluster the population along a number of characteristics, but essentially using the similarity of their capability sets (welfare distribution). \\
For the rest of the paper we will focus essentially in presenting the first step in details.

\subsubsection{General presentation of the model}

In order to model citizens capabilities, we will use Mathematical Programming
(MP). It is a tool ordinarily used for decision aiding purposes. It is both efficient (thanks to the availability of different solvers), easily understandable and systemic.

Moreover, it allows constraints based modeling. Capabilities are the set of the different \emph{combinations of functionings} that are \emph{feasible} given certain resources and individuals' attributes. Those \emph{resources} are easy to model in a MP. Besides, using MP allows to represent functionings thanks to decision variables, while solving the model enables establishing ``optimal'' \emph{combinations of functionings}. Nonetheless, as far as we know, there have been no attempts to use MP in order to represent citizens' capabilities.

%\begin{note}
 % We need to justify better why we use Mathematical Programming. The above
  %reasons are reasonable, but not sufficient.
%\end{note}

%What we try to do is, given a citizen private resources, access to Commons,
%conversion factors and values, to represent the individual capability set.
The model will focus on the Doings of the citizen (our \emph{decisions}
variables), which will be constrained by citizen's private resources and access to Commons and his/her \emph{conversion factors}. The Beings of a citizen will be derived from his/her Doings through his/her conversion factors and values. The idea is to find the Pareto set of points in the Beings spaces. The set of ``interesting beings'' and their associated doings is going to be considered as our capability set. Having this set of points, the idea is to simulate an impact upon a Common, to create the new possible Pareto dominant set of Beings, and study the difference between the possible Doings and Beings of a citizen.

We need to define some of the terms that we are going to use and redefine some notions of the Capability approach that may differ in our generic model. For the sake of clarity, parameters are given supposing a linear problem within MP.

\begin{itemize}
    \item \emph{Private resources}:\\
    \begin{equation*}R^i =
\begin{bmatrix}
r^i_{1} & r^i_{2} & \cdots & r^i_{l}
\end{bmatrix}\\
\end{equation*}
Each citizen $i$ owns/has access to different sets of private resources. A resource
        vector can be composed by commodities, such as the amount of money
        and simple private goods that citizens possess. It can also contain
        more complex resources such as time or energy. $R^i$ is a vector
        containing $l$ different type of private resources.
    \item \emph{Commons}:     \begin{equation*}C =
\begin{bmatrix}
c_{1} & c_{2} & \cdots & c_{k}
\end{bmatrix}
\end{equation*}\\
We consider the definition given by
        \cite{ostrom1990governing}: a Common is a good that is a natural or
        man-made resource, non-excludable (\emph{i.e. it is not possible to
        exclude people from its utilisation}) and rivalous (\emph{i.e. the
        consumption or utilisation of the good by an individual will have
        impact on others individuals utilisation, by reducing the quantity of good available or by decreasing the 'utility' derived from the
        utilisation of the good}). A Commons vector will be a set of $k$ Commons that a community has access to.
    \item \emph{Doings}
        \begin{equation*}X =
\begin{bmatrix}
x_{1} & x_{2} & \cdots & x_{j}
\end{bmatrix}\\
\end{equation*}
    Our decisions variables. These are actions or activities that citizens
        can do, in order to ``achieve'', ``obtain'' or ``reach'' one or more goals. More precisely, we consider a \emph{Doings} $\overline{X}^z$ as a combination of actions/activities that an individual can do at the same ``time''. For instance, two possible Doings for an afternoon can be; $\overline{X}^1$ going to run, see some friends and eat healthy \emph{or} $\overline{X}^2$ play video-games and eat unhealthy. We consider that functionings are constructions of doings. The \emph{Doings set} is all the possible Doings that a citizen can decide to achieve; all the different combinations of actions that a citizen can perform in a given interval of time.
    \item \emph{Beings vector (noted $B^z$):} is the image of our variables
        (\emph{Doings}) in the welfare dimension space. A \emph{being}
        $b^z_h$ is the ``quantity'' of how ``well'' a citizen is on a given
        \emph{welfare dimension} $h$ and Doings $X^z$. For instance, being
        healthy or unhealthy are two beings of the welfare dimension of
        health. As for Doings, a Beings $B^z$ is a combination of beings of a citizen in a certain functioning $X^z$, for instance, $B^1$ being
        happy and healthy \emph{or} $B^2$ being happy and unhealthy. The
        \emph{Beings set} is all the possible Beings that a person can decide to be.
    \item \emph{Conversion factors:} As described by
        \cite{robeyns2003capability} and previously in this paper (in section \ref{CA}). They are the personal, social and environmental conversion factors, that convert resources (private or Common) into
        functionings. They both influence the \emph{conversion} and the
        \emph{transformation} matrix.
    \item \emph{Values:} They are citizen's conception of what is good. They come from personal history and psychology of the citizen. Of course, values interact with personal, social and environmental factors, they both influence each other, but two citizens with the same conversions factors do not have necessarily have the same values. Values are purely subjective. They influence the Transformation Matrix.
    \item \emph{Conversion Matrix}:  \begin{equation*} A^i =
\begin{bmatrix}
a^i_{1,1} & \cdots & a^i_{1,l}  & a^i_{1,l + 1}& \cdots & a^i_{1,l+k} \\
a^i_{2,1} & \cdots & a^i_{2,l}  & a^i_{2,l +1}& \cdots & a^i_{2,l+k} \\
\vdots  & \ddots &\vdots   & \vdots& \ddots & \vdots  \\
a^i_{j,1} & \cdots & a^i_{j,l}  & a^i_{j,l + k}& \cdots & a^i_{j, l+k}
\end{bmatrix}
\end{equation*}

Determines how many private resources and Commons a citizen will consume/earn for a set of Doings. For instance, going to university will use one hour of a citizen's time, 45 minutes of public transport and some energy and money. The consumption of resources and Commons depend on \emph{personal conversion factors} (how fast I walk?); \emph{social         conversion factors} (am I allowed to use public transport?); and         \emph{environment conversion factors} (where do I live?). The         consumption (or earning) of a resource $l$ (resp Common $k$) for a         Doings $\overline{X}^z$ is determine by $\Phi^i_l(\overline{X}^z, A^i_l)$ (resp $\Phi^i_k(\overline{X}^z, A^i_{l+k})$).

    \item \emph{Transformation matrix}:
    \\ \begin{equation*}
W^i =
\begin{bmatrix}
w^i_{1,1} & w^i_{1,2} & \cdots & w^i_{1,h} \\
w^i_{2,1} & w^i_{2,2} & \cdots & w^i_{2,h} \\
\vdots  & \vdots  & \ddots & \vdots  \\
w^i_{j,1} & w^i_{j,2} & \cdots & w^i_{j,h}
\end{bmatrix}
\end{equation*}

Determines how a Doings $\overline{X}^z$ will be transformed into a Beings $B^z $. These are influenced by conversion factors and values. Two citizens
can have exactly the same Doings set, but different Beings set. The         impact of an action on an individual ``is'' be influenced by personal, social and environmental factors. In addition, one can derive a strong positive impact on a welfare dimension such as \emph{happiness} and \emph{living in respect with his/her (environmental) value} from going to work on bike, while another can derive low or negative impact on these beings dimensions for the same action. The level of being of $i$ on the welfare dimension $h$ is calculated by $f^i_h(\overline{X}^z, W^i_h)$.

    \item \emph{Welfare representation:} This is the set of Beings that are
        not Pareto dominated in the Beings set. They are considered as the
        ``interesting'' Doings that an individual can achieve. The welfare
        representation is noted $\mathcal{W}(\cdot)$.
\end{itemize}

Then, given a citizen $i$, our generic model for the welfare representation is:

\begin{table}[H]
    \centering
\begin{tabular}{lllll}
     $\max_{\forall h}$& $\langle f_h(X, W^i_h)\rangle$ \\
      $s.t.$ &  &\\
      & $\Phi^i_{l}(X, A^i_l)$ & $\leq$ & $R^i_l$ & $\forall l$\\
      & $\Phi^i_{k}(X, A^i_{l+k})$ & $\leq$ & $C_{k}$ & $\forall k$\\
\end{tabular}
\end{table}{}

 % \max_{\forall h}\langle f_h(X, W^i_h)\rangle  \\
 % s.t.\\
  %\forall l:   \Phi^i_{l}(A_{l}, X)\leq A_l \\
  %\forall k:   \Phi^i_{k}(A_{l+k}, X)\leq C_k \\

%\begin{note}
%generic model should like that: \\
%  $\max_{\forall l}\langle f^{\chi}_l(w_{jl},x_j)\rangle $ \\
%  subject to \\
%  $\forall i \Phi_{i\chi}(a_{ij},x_j)\leq\{b_i,c_i\}$ \\
%If this is correct, then I need to see the above bullets explained with respect
%to this generic model. I suppose this is the only way to agree upon what we
%mean with them (and why ...)
%\end{note}

\begin{sidewaysfigure}[]
    \includegraphics[scale=0.48]{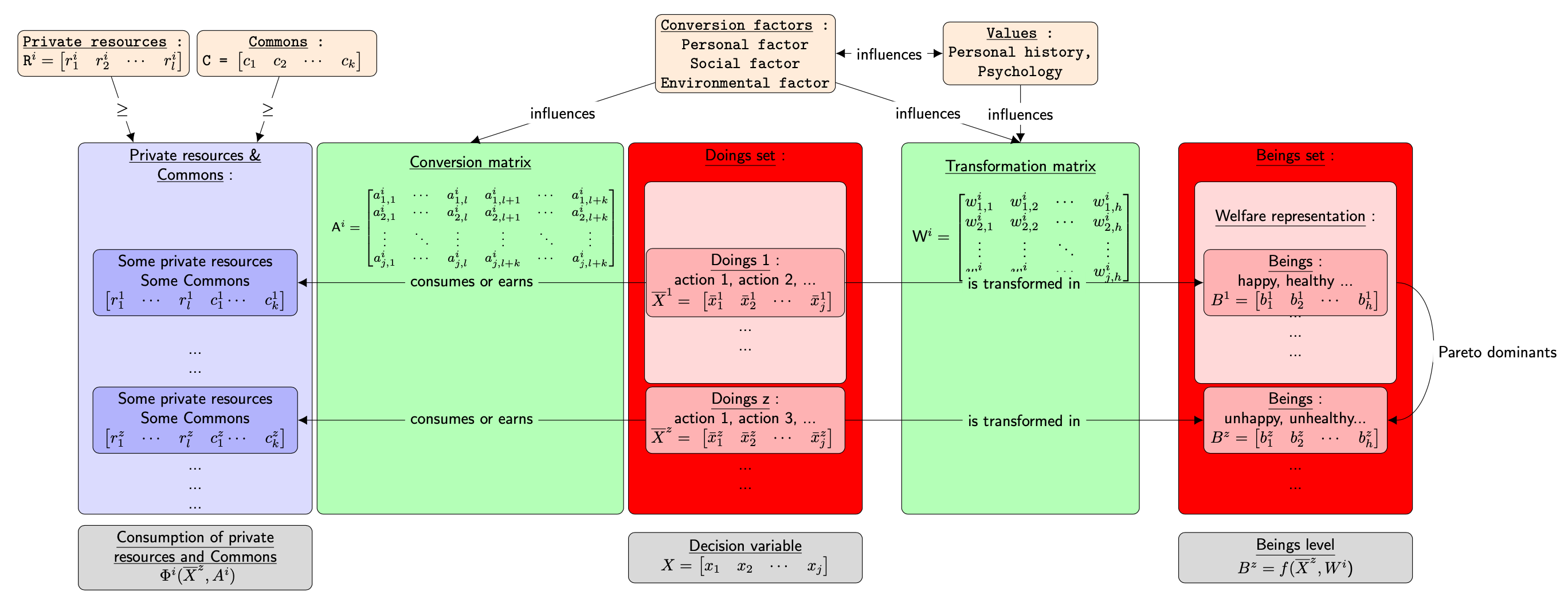}
    \caption{A schematic representation of the model}
    \label{fig:LandscapeFigure}
\end{sidewaysfigure}

A schematic representation of how the different elements of the model interact is proposed in figure \ref{fig:LandscapeFigure}. Personal psychology and history are interacting with the conversion factors. The personal utilisation functions are derived from the conversion factors, and given the citizen private resources and access to Commons we can find the Doings set. Then the values and utilisation functions convert Doings into Beings. The transformation of the Doings set gives the Beings set. From a more MP perspective: private resources and Commons are consumed to achieved some Beings level through the realisation of some Doings.

If citizen (1) and (2) have the same private resources and Commons, they may
not have the same Doings set because of different utilisation function.

If (1) and (2) have the same private resources, Commons and utilisation
function, then they have the same Doings set but not necessarily the same
Beings set because of different values.

Note that an \emph{action} has to be distinguished from its finality. Two
actions can have the same goal or result, but different consumptions of
resources and Commons as well as having different impacts on beings. For
instance, we have to make the distinction between going to work using public
transports and going to work by car. Even if the finality is the same, the
consumption of resources and Commons will be different as well as the Beings of the individual. Moreover a citizen can choose an action, (1) to be able to
achieve another action, or (2) because of his/her impact on its Beings (or
both). For example, someone can (1) take the subway to go to the center because (2) s.he wants to go to the museum.

%\begin{note}
%  I downsized figure 2.1, but consider the option of putting it in landscape
%  mode.
%\end{note}

\subsubsection{Simple example}

Here, we propose a simple example in order to illustrate the different concepts proposed previously.

In this example we consider a city with 5 intersections between streets, streets being considered as \emph{Commons} (Figure \ref{fig:ex1}). A citizen is living in $1$ and wants to optimize his/her Beings, that is composed of two \emph{welfare dimensions}: Beauty and Health. To maximise his/her Beings, the citizen can only walk through the streets of the city. Each street will have an impact on his/her being (in red in the Figure \ref{fig:ex1}), it is the \emph{transformation matrix} $W^1$. Taking a street will also consume \emph{private resources}: Energy and Time, (in blue in the Figure \ref{fig:ex1}), as well as ``one utilisation of the street'', through the \emph{conversion matrix} $A^1$. The citizen starts with a quantity of energy $=10$ and quantity of time $=10$, his/her \emph{resource vector} $R^1$. The Commons vector $C$ is the set of streets. To make the example easier to understand, we will consider that a citizen can only use a street once. The last constraint is that the citizen has to come back to his/her home at the end of the walk. The parameters of our problem are then:
    \\ \begin{equation*}
R^1 =
\begin{bmatrix}
10 & 10\\
\end{bmatrix}
\end{equation*}
\begin{equation*}
C =
\begin{bmatrix}
1 & 1 &1&1&1&1&1&1\\
\end{bmatrix}
\end{equation*}
\begin{equation*}
A^1 = \begin{bmatrix}
2 & 3 &1&1&5&2&2&5\\
3 & 2& 1& 1&5&2&2&5\\
1 & 1 &1&1&1&1&1&1\\
1 & 1 &1&1&1&1&1&1\\
1 & 1 &1&1&1&1&1&1\\
1 & 1 &1&1&1&1&1&1\\
1 & 1 &1&1&1&1&1&1\\
1 & 1 &1&1&1&1&1&1\\
1 & 1 &1&1&1&1&1&1\\
1 & 1 &1&1&1&1&1&1\\
\end{bmatrix}
\end{equation*}
\begin{equation*}
W^1 =
\begin{bmatrix}
 3 & 0 &0&2&2&4&2&1\\
1 & 3& 2& 1&2&1&3&1
\end{bmatrix}
\end{equation*}
\begin{equation*}
x =
\begin{bmatrix}
 x_{12} & x_{13} &x_{14}&x_{15}&x_{23}
 &x_{24}&x_{35}&x_{25}\\

\end{bmatrix}
\end{equation*}

In this example, there are only $16$ possible actions: walking in the street
$12$ (from vertex 1 to vertex 2), $13$, $14$, $15$, $23$, $24$, $35$ or $45$, and their reverse ($21$, $31$, $41$, $51$, $32$, $42$, $53$ or $54$). They
constitute our decisions variable (in our model we make no distinction between a street and its reverse so we only have 8 actions). There are only nine possible paths which consume less or equal than Energy$=10$ and Time$=10$, or in other words, considering the reverse paths as equivalent there are only five different Doings in the Doings set $=\{\overline{X}^A, \overline{X}^B, \overline{X}^C, \overline{X}^D, \overline{X}^E\}$:

\begin{itemize}
    \item $\overline{X}^A$: Doings composed of actions $12$, $23$ and $31$
        (or $13$, $32$ and $21$) which will consume (Energy: 10, Time: 10)
        and will have an impact of (Beauty: 6, Health: 6) on Beings.
    \item $\overline{X}^B$: Doings composed of actions $13$, $35$ and $51$
        (or $15$, $53$ and $31$) which will consume (Energy: 6, Time: 5) and
        will have an impact of (Beauty: 4, Health: 7) on Beings.
    \item $\overline{X}^C$: Doings composed of actions $12$, $24$ and $41$
        (or $14$, $42$ and $21$) which will consume (Energy: 5, Time: 6) and
        will have an impact of (Beauty: 5, Health: 6) on Beings.
    \item $\overline{X}^D$: Doings composed of actions $14$, $45$, $51$ (or
        $15$, $54$ and $41$) which will consume (Energy: 7, Time: 7) and will
        have an impact of (Beauty: 3, Health: 4) on Beings.
    \item $\overline{X}^E$: Staying at home (no actions) which will consume
        (Energy: 0, Time: 0) and will have an impact of (Beauty: 0, Health:
        0) on Beings.
\end{itemize}

\begin{figure}[H]
\begin{tikzpicture}[scale=0.85]
\begin{axis}[
    title={},
    xlabel={Energy},
    ylabel={Time},
    xmin=0, xmax=12,
    ymin=0, ymax=12,
    xtick={0,2,4,6,8,10},
    ytick={0,2,4,6,8,10},
    grid=major,
    axis x line=bottom, axis y line = left,
]
 \addplot[
    color=blue,
    mark=square,
    ]
    coordinates {
    (10,10)
    };

\addplot[
    color=blue,
    mark=square,
    ]
    coordinates {
    (6,5)
    };
 \addplot[
    color=blue,
    mark=square,
    ]
    coordinates {
    (5,6)
    };
     \addplot[
    color=blue,
    mark=square,
    ]
    coordinates {
    (7,7)
    };
         \addplot[
    color=blue,
    mark=square,
    ]
    coordinates {
    (0,0)
    };
\node [above right,color=blue] at (100,100) {$\overline{X}^A$};
\node [above right,color=blue] at (60,50) {$\overline{X}^B$};
\node [above right,color=blue] at (50,60) {$\overline{X}^C$};
\node [above right,color=blue] at (70,70) {$\overline{X}^D$};
\node [above right,color=blue] at (0,0) {$\overline{X}^E$};
\addplot[color= green,
]
coordinates {(10,0) (10,10)};
\addplot[color= green,
]
coordinates {(0,10) (10,10)};
\end{axis}
\end{tikzpicture}
    \centering
    \caption{Consumption of resources of the citizen regarding the different Doings}
    \label{fig:const1}
\end{figure}
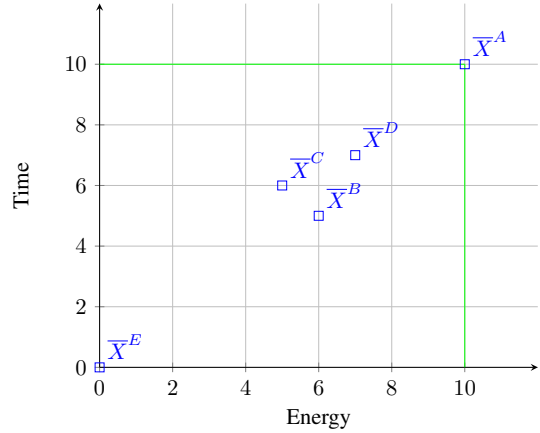

Figures \ref{fig:const1} and \ref{fig:im1} are the graphical representation of the Doings set respectively on the space of private resources and in the Beings space (\emph{i.e. the graphical representation of the Beings set}).

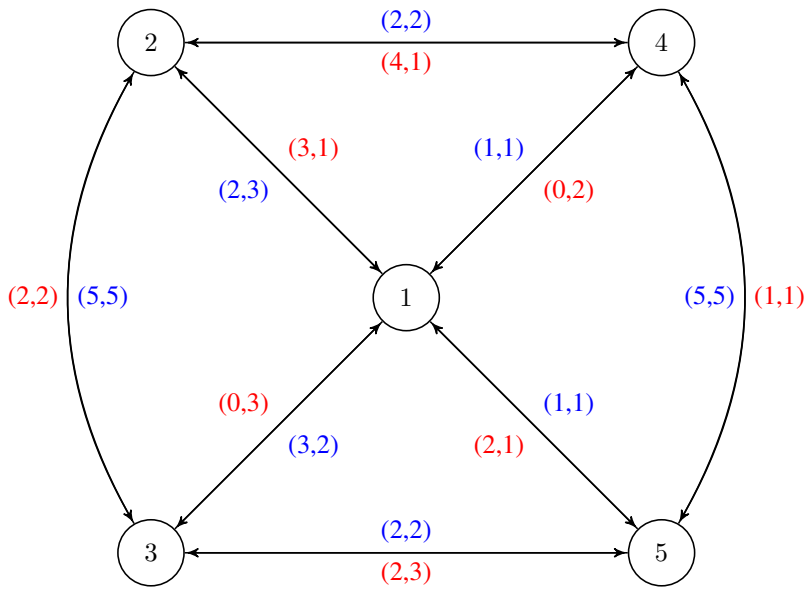
\begin{figure}[H]
\begin{tikzpicture}[->,>=stealth',shorten >=1pt,auto,node distance=4.8cm,
                    semithick][H]
  \tikzstyle{every state}=[fill=none,draw=black,text=black]
  \node[state] (1)                    {$1$};
  \node[state] (2) [above left of=1] {$2$};
\node[state] (3) [below left of=1] {$3$};
\node[state] (4) [above right of=1] {$4$};
\node[state] (5) [below right of=1] {$5$};
  \path (1) edge     []         node {\textcolor{blue}{(2,3)}} (2);
  \path (1) edge     []         node {\textcolor{blue}{(3,2)}} (3);
  \path (1) edge     []         node {\textcolor{blue}{(1,1)}} (4);
  \path (1) edge     []         node {\textcolor{blue}{(1,1)}} (5);
  \path (2) edge     []         node {\textcolor{red}{(3,1)}} (1);
  \path (2) edge     [bend right]         node {\textcolor{blue}{(5,5)}} (3);
  \path (2) edge     []         node {\textcolor{blue}{(2,2)}} (4);
  \path (3) edge     []         node {\textcolor{red}{(0,3)}} (1);
  \path (3) edge     [bend left]         node {\textcolor{red}{(2,2)}} (2);
  \path (3) edge     []         node {\textcolor{blue}{(2,2)}} (5);
  \path (4) edge     []         node {\textcolor{red}{(0,2)}} (1);
  \path (4) edge     []         node {\textcolor{red}{(4,1)}} (2);
  \path (4) edge     [bend left]         node {\textcolor{red}{(1,1)}} (5);
  \path (5) edge     []         node {\textcolor{red}{(2,1)}} (1);
  \path (5) edge     []         node {\textcolor{red}{(2,3)}} (3);
 \path (5) edge     [bend right]         node {\textcolor{blue}{(5,5)}} (4);
\end{tikzpicture}
   \caption{Cost \textcolor{blue}{(Energy, Time)} and  impact on welfare
   \textcolor{red}{(Beauty, Health)} associated to streets of a city for a citizen}
   \label{fig:ex1}
\end{figure}
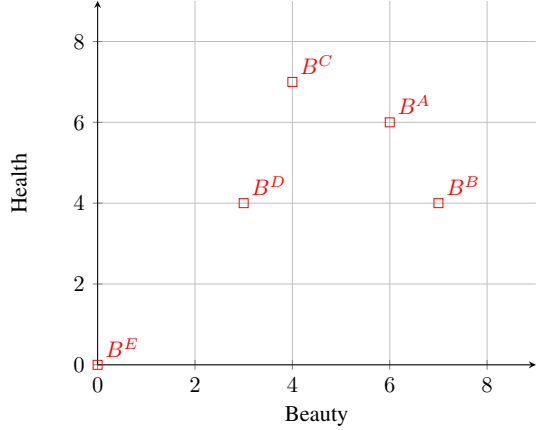
\begin{figure}
\begin{tikzpicture}[scale=0.85]
\begin{axis}[
    title={},
    xlabel={Beauty},
    ylabel={Health},
    xmin=0, xmax=9,
    ymin=0, ymax=9,
    xtick={0,2,4,6,8},
    ytick={0,2,4,6,8},
    grid=major,
    axis x line=bottom, axis y line = left,
    %legend pos=north west,
    %ymajorgrids=true,
    %grid style=dashed,
]
 \addplot[
    color=red,
    mark=square,
    ]
    coordinates {
    (6,6)
    };

\addplot[
    color=red,
    mark=square,
    ]
    coordinates {
    (7,4)
    };
    %\legend{2}
 \addplot[
    color=red,
    mark=square,
    ]
    coordinates {
    (4,7)
    };
     \addplot[
    color=red,
    mark=square,
    ]
    coordinates {
    (3,4)
    };
     \addplot[
    color=red,
    mark=square,
    ]
    coordinates {
    (0,0)
    };
\node [above right,color=red] at (600,600) {$B^A$};
\node [above right,color=red] at (700,400) {$B^B$};
\node [above right,color=red] at (400,700) {$B^C$};
\node [above right,color=red] at (300,400) {$B^D$};
\node [above right,color=red] at (0,0) {$B^E$};
\end{axis}
\end{tikzpicture}
    \centering
    \caption{Beings set of the citizen}
    \label{fig:im1}
\end{figure}

What a citizen can do is constrained by his/her private resources, in green in Figure \ref{fig:const1}, and the street network. Every Doing that is contained in the square drawn by the axis and the greens line is a realisable Doing. It is clear that the Doings $\overline{X}^D$ and $\overline{X}^E$ will never be chosen by the citizen, as their associated Beings are Pareto dominated respectively by $B^A, B^B, B^C$ and $B^A, B^B, B^C, B^D$. The Beings set containing the values of the citizen, s.he will necessarily choose a solution in the his/her welfare representation ($\{B^A, B^B, B^C\}$).

\subsection{Formal presentation of the model}

%\subsubsection{Model}

%\begin{note}
%  Simplify ...
%\end{note}

%\paragraph{Indices}

%\begin{tabular}{lcl}
%         $h$ & $=$ &  Dimension of beings; \{$h=1$ Felling safe,$h=2$ Having a
%         good social life, $h=3$ Being healthy,\\&& $h=4$ Being happy ...$h = H$\}; \\
%         $i$ & $=$ & Citizen; $\{i=1, i=2, ..., i=I\}$ \\
%         $j$ & $=$ & Action; $\{j = 1$ Play football, $j = 2$ Work, $j=3$ See the doctor, ... $j = J$ $\}$ \\
 %        $l$ & $=$ & Resources; $\{l = 1$ Money, $l = 2$ Time, $l = 3$ Car, ..., $l = L\}$ \\
  %       $k$ & $=$ & Commons; $\{ k = 1$ Road, $k = 2$ Forest, $k = 3$ River..., $k = K \}$
%\end{tabular}

%\paragraph{Given data}

%\begin{tabular}{lcl}
 %        $A^i_{l}$ & $=$ & Quantity of resources $l$ that citizen $i$ possesse.\\
  %       $A^i_{jl}$ & $=$ & The quantity of ressource $l$ that is consumed
   %      ($< 0$) or gained ($> 0$)\\&& by $i$ for doing the $j$. \\
    %     $C_k$ & $=$ & Quantity of commons $k$ available;\\
     %    $C^i_{jk}$ &$=$&The quantity of commons $k$ that is consumed by citizen $i$ for doing $j$\\
      %   $\Delta_k$ & $=$ & The general appropriation of the commons $k$\\
       %  && (\emph{the quantity of the commons $k$ that the society is consuming});\\
        % $W^i_{jh}$ & $=$ & The transformation of the doing $j$ on the beings $h$ of citizen $i$.\\
%\end{tabular} \\

%$A^i_{jl}$ and  $C^i_{jk}$ are conversions parameters. $W^i_{jh}$ are
%transformation parameters.
As already mentioned, designing a public policy implies being able to:\\
- observe the present situation and distribution of welfare;\\
- anticipate the impact of doing nothing;\\
- anticipate the impact of implementing a policy.

We note $\mathcal{X}_i = \{A^i, C, R^i, W^i\}$ the citizen's \emph{present
state}, at the time of the observation. We consider that his/her present state is represented by the welfare representation $\mathcal{W}(\mathcal{X}_i)$ at the present time (before any Commons' damage or new policy). Then, we need to be able to anticipate the impact of doing nothing and implementing new policies. We will focus on the impact of different scenarios of Commons' damage and public policy. These scenarios will be seen as modification of parameters from the \emph{present state}. Note that for the time being the modelling of likelihoods for such threats is not considered. This is not to underestimate the importance of this topic, but only the result of our focus upon welfare representation and measurement.

The idea is to find the citizens welfare representation in a certain scenario
in order to have a mesure of the intensity of a hazard or the impact of a
public policy. The welfare representation of citizen $i$ in the case of
scenario $s$ is noted $\mathcal{W}(\mathcal{X}_i|scenario_s)$ or
$\mathcal{W}(^s\mathcal{X}_i)$.

\paragraph{Variables}

Our mathematical optimisation program is a mix-variable problem. We use binary variables ($x_j \in \{0, 1\}$) when the issue is whether a certain ressource or a common are used (can be used) or not. We use real valued variables ($x_{j'}\in \mathcal{R}$) when ressources or commons are consumed (at different possible levels).

\paragraph{Constraints over private resources}

The consumption of a private resource $l$ by a citizen to achieve his/her
Doings should be inferior to the quantity of resources $R_l^i$. The constraint over the quantity of the private resources is not necessarily linear and is calculated through the function $\Phi^i_l(X, A^i_l = \begin{bmatrix} a^i_{1,1}
& \cdots & a^i_{j,l}
\end{bmatrix})$. The quantity of private resources consumed by the citizen
should be inferior to the quantity of resources that s.he possesses:
 \begin{tabular}{lclr}
     $\Phi^i_l(X, A^i_l)$ & $\leq$ & $R^i_l$ & for all $l$.
\end{tabular}\\
Supposing the problem being linear we get:
 \begin{tabular}{lclr}
     $X \cdot A^i_l$ & $\leq$ & $R^i_l$ & for all $l$.
\end{tabular}\\
%\paragraph{Variables}
%\begin{tabular}{lcl}
%     $x^i_j$& $=$ & number of times the citizen $i$ chooses action $j$,\\
%\end{tabular}\\

Public policies can have a impact both on private resources ($R^i$) and
conversion matrix ($A^i$). For instance a new tax will impact $r^i_{money}$ and a road tool an impact $a^i_{highway, money}$. Then, the constraints over
private resources in a scenario $s$ are of the form:

\begin{tabular}{lclr}
     $^s\Phi^i_l(X, ^sA^i_l)$ & $\leq$ & $^sR^i_l$ & for all $l$.
\end{tabular}\\

%A citizen can choose an action, (1) to be able to achieve another action, or
%(2) because of his/her impact on its Beings (or both). For example, someone can
%(1) take the subway to go to the center because (2) s.he wants to go to the
%museum.

\paragraph{Constraints over Commons}

We will distinguish two types of Commons; those that are utilised and those
that are consumed. In the first case, the ``quantity'' of Commons is not
reduced when a citizen is using it, but it reduces the utility derived for
other citizens. For instance, a road can be considered as an ''utilised''
Common. More people on a road will not consume it, but it will create traffic
jam, decreasing the utility of the road. For those sort of Commons, we can use binary representation; $C_k \in \{0, 1\}$, the Common either exist or not. All actions $j$ either use this Common $k$, ($a^i_{j, l+k} = \epsilon$ with epsilon a very small value), or don't use it ($a^i_{j, l+k} = 0$). If the Common becomes not available ($C_k = 0$), because of a damage, all actions using $k$ ($\exists a^i_{j, l+k} = \epsilon$) are then impossible to do and set to 0. For such types of Commons, the consumption of other citizens have a great impact on a particular citizen conversion matrix. For instance, the damage of a particular road can influence the traffic on the entire road network. The use of different Commons can vary regarding different scenarios, then we have the following constraints:

 \begin{tabular}{lclr}
     ${}^s\Phi^i_k(X, {}^sA^i_{l+k})$ & $\leq$ & ${}^sC_k$ & for all $l$.
\end{tabular}

In the second case Commons are consumed. For instance, if a citizen consumes a quantity of water from a groundwater, this can not be consumed by someone else. In this situation, the Common  pool resources which can be consumed by a particular citizen is what is left by the other citizens. The general
consumption of a good $k$ in a scenario $s$ is noted $^s\Delta_k$. Consumable
Commons use continuous representations; $C_k, \Delta_k \in \mathcal{R}$. Given different Commons, the quantity left can also affect the conversions
parameters; it is more difficult to catch a fish if there is little left in a
lake than if there are many. For consumable Commons, we have the following
constraints:

 \begin{tabular}{lclr}
     $^s\Phi^i_k(X, ^sA^i_{l+k})$ & $\leq$ & ${}^s C_k -{}^s  \Delta_k$ & for all $l$.
 \end{tabular}\\
%$\Psi^i_k(C^i_{1k} x_1^i, C^i_{2k} x_2^i, ..., C^i_{Jk} x_J^i)$ and is not
%necessarily linear. It should be inferior to the quantity of Commons that
%haven't been consumed by the rest of citizens, $ C_k - \Delta_K$. We have:

 %\begin{tabular}{lclr}
  %   $\Psi^i_k(C^i_{1k} x_1^i, C^i_{2k} x_2^i, ..., C^i_{Jk} x_J^i)$ & $\leq$ & $ C_k - \Delta_K$ & for all $i$ and $k$
%\end{tabular}

%\begin{note}
 % We need to remember that we have two different types of ``commons'' (seen in
  %terms of resources). \\
  %The first type are resources which either exist or not (a bridge, an access
  %etc.). For these although exists a law of consumption (seen in terms of
  %maintenance) we need to insert constraints which exclude certain actions as
  %soon these resources turn to be 0. \\
  %The second type are resources which are consumed (water ...). For these we
  %need to construct a consumption model depending upon the number of individual
  %who consume the good ... \\
  %We need to think better ...
%\end{note}

\paragraph{Objective Function}

The objective for citizen $i$ is to ``maximise'' his or her welfare,
in other terms in order to compute the Welfare representation
$\mathcal{W}(\mathcal{X}_i)$, we need to compute the Pareto-frontier of the
Beings set for citizen $i$.

If we consider all dimensions of welfare that could affect the citizens' choice, we assume that all individuals would/should choose a solution that is Pareto efficient. For instance, considering the Sen's example of a fasting person for political reason, focusing only on two functioning such as \emph{being able to obtain an adequate amount of food} and \emph{being able to be in good heath} is not sufficient to fully understand the citizens' welfare and act. Indeed, fasting is not a solution in the Pareto frontier, because by eating, the citizen should be able to be better in both dimensions. But considering more dimensions of welfare such as \emph{the expression of his/her political opinion}, this person is choosing a solution in the Pareto frontier. In the context of decision aiding, and broadly from an operational point of view, it is clearly impossible to consider all welfare dimensions that could affect a citizens' choice. ``Simply'' computing the Pareto front is  therefore an approximation.

The transformation of actions in different beings are not necessarily
independent. For example, going to swim, going cycling or going running may
have a good impact on citizen's health, but for a non-athletic person, doing
the 3 activities the same day may have a negative impact on his/her health (as s.he may be injured). Then, the ``quantity'' of the beings $h$ is converted through the non necessarily linear function $F_h(X, W^i_h)$.

For one Welfare representation, we get:
\begin{tabular}{lllr}
     $\forall h$ $\max F_h(X, W^i_h)$& \\
\end{tabular}

As for other parameters, the transformation matrix can change regarding the
scenario. For instance, the comfort in a subway is lower when it is full than
usually. Then, for a scenario $s$, we have:
\begin{tabular}{lllr}
     $\forall h$ $\max ^sF_h(X, ^sW^i_h)$& \\
\end{tabular}

For the time being, $W^i$ is considered as being given. For an empirical use of the model, the transformation matrix $W^i$ needs to be learned. The reader can note that one way to learn $W^i$ is to use subjective value measurement
procedures \citep{Raiffa1969}. If we consider that each welfare dimension is
independent, the transformation matrix can be learned using multi-attribute
value theory \citep{keeney1993decisions,winterfeldt}. We will not discuss the
use of such methods in this paper.

%\textbf{The citation of Rescher as far as learning values and utilities is not
%appropriate. The whole paragraph may need to me moved elsewhere in the text.}

\paragraph{Clustering  and comparison} To be able to use the framework in a
process of decision aiding, we need to be able to find clusters of citizens
sharing same values and welfare representation. We also need to be able to find a way to compare different policies from the point of view of (cluster of) individuals. The clustering and comparison problem \citep{zeumo2014new,Fancellolurning,FancelloetalSEPS2020} are not considered in this work, but will be investigated in the future .

Essentially there are two ways to compare citizens: one comparing their Pareto frontiers (their capability sets), the other comparing the value functions through which the Pareto frontiers have been constructed. One citizens are clustered (and thus, become targets of policies) we can consider the design of ad-hoc policies taking into account the specific characteristics of these clusters.

%And the general model of an individual $i$ is  then:

%\begin{tabular}{llllllllll}
%     $\max$ & $F_1($ & $W^i_{1,1} x^i_{1}$ &  $W^i_{2,1} x^i_{2}$ & ...  & $W^i_{J,1} x^i_{J})$  \\
%     &&&...&&&\\
%     $\max$ & $F_h($ & $W^i_{1,h} x^i_{1}$ &  $W^i_{2,h} x^i_{2}$ & ...  & $W^i_{J,h} x^i_{J})$  \\
%     &&&...&&&\\
%     $\max$ & $F_H($ & $W^i_{1,H} x^i_{1}$ &  $W^i_{2,H} x^i_{2}$ & ...  & $W^i_{J,H} x^i_{J})$  \\
%     \\
%    $s.t.$ &$\Phi^i_1($&$ A^i_{1,1} x_1^i$ & $A^i_{2,1} x_2^i$ & ... & $A^i_{J,1} x_J^i)$ & $\leq$ & $A^i_1$\\
%    &&&...&&&\\
%    &$\Phi^i_l($& $A^i_{1,l} x_1^i$ & $A^i_{2,l} x_2^i$ & ... & $A^i_{J,l} x_J^i)$ & $\leq$ & $A^i_l$\\
%    &&&...&&&\\
%    &$\Phi^i_l($&$ A^i_{1,L} x_1^i$ & $A^i_{2,L} x_2^i$ & ... & $A^i_{J,L} x_J^i)$ & $\leq$ & $A^i_L$\\
%    \\
%    &$\Psi^i_1($&$ C^i_{11} x_1^i$ & $C^i_{21} x_2^i$ & ...& $C^i_{J1} x_J^i)$ & $\leq$ & $ C_1 - \Delta_1$\\
%    &&&...&&&\\
%    &$\Psi^i_k($&$ C^i_{1k} x_1^i$ & $C^i_{2k} x_2^i$ & ...& $C^i_{Jk} x_J^i)$ & $\leq$ & $ C_k - \Delta_k$\\
%    &&&...&&&\\
%    &$\Psi^i_K($&$ C^i_{1K} x_1^i$ & $C^i_{2K} x_2^i$ & ...& $C^i_{JK} x_J^i)$ & $\leq$ & $ C_K - \Delta_K$\\
%    &&&&&$x^i_j$ & $\in$ & $N$
%\end{tabular}\\

\subsubsection{Example}

\begin{figure}
    \centering
    \includegraphics[width=\textwidth]{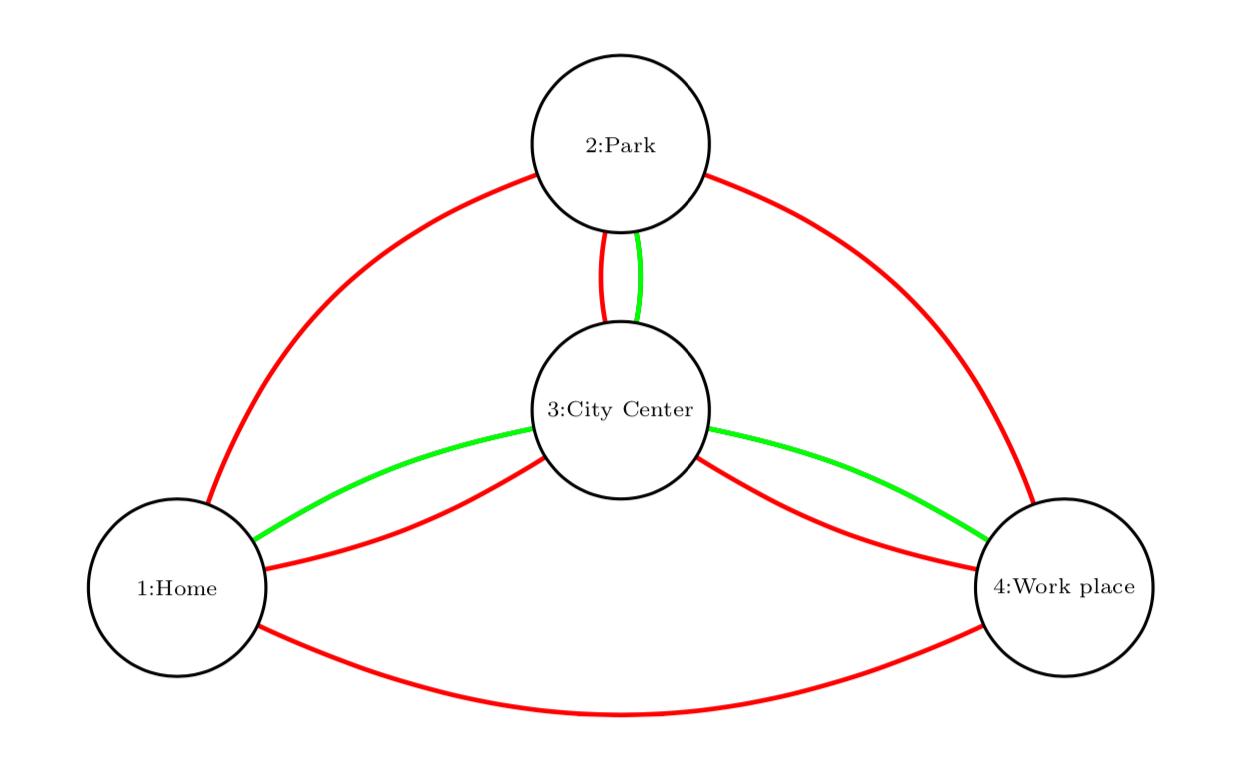}

\caption{Graph of the city of citizens $1$ and $2$, the arcs in green
represent public transports and arcs in red represent roads.}
    \label{fig:graph}
\end{figure}

This example is constructed in order to demonstrate how we want to use our
framework in the context of collective threats. We want to represent the
welfare of two individuals: $1$ and $2$, that are in the current states
$\mathcal{X}_1$ and $\mathcal{X}_2$. They are living in the same city, and have the same representation of it, through the graph $G(V,E)$ Figure
\ref{fig:graph}:

\begin{itemize}
    \item V: The set of vertices, that represent points of opportunities
        (\emph{i.e. 1: Home, 2: Park, 3: City center, 4: Work place}). The
        two citizens are living in the same neighbourhood $1$, and are both
        working in $4$. On each vertex, citizens can perform actions, that
        can be derived in beings through the transformation matrix $W^i$ (see table \ref{fig:tab1} and \ref{fig:tab2}). Those transformations parameters are personal for each citizen, and depend both on the citizens' conversion factors and values. An action consumes private resources and Commons $A^i$ (see table \ref{fig:tab1} and \ref{fig:tab2}), these consumptions are different regarding only the conversion factors.
    \item E: The set of road $e_{vwr}$ and public transport (PT) $e_{vwp}$
        that links two vertices $vw$. Both are considered as Commons. To make the example simpler, we consider that the transformation into beings of the utilisation of a road (resp. PT) and the consumption of private resources are the same regardless of the road (resp. PT) the citizen uses (see table \ref{fig:tab1} and \ref{fig:tab2}). The
        utilisation of these edges are considered as actions.
\end{itemize}

Each citizen has a quantity of private resources $A^i =\{A^i_1 = $ Money, $
A^i_2 = $ Time, $  A^i_3 = $ Car $\}$, described in table
\ref{fig:commodities}. The goal of a citizen is to maximise his/her Beings ($f^{i}_h =$ Health,$f^{i}_p =$
Pleasure)
through his/her transformation functions. The solutions obtained will establish the Pareto frontier
representing the citizen $i$; the Welfare representation
$\mathcal{W}(\mathcal{X}_i)$.

    % Flush earlier floats (otherwise order might not be correct)
% empty page style (?)
    \begin{landscape}

    % Landscape page

\begin{table}[H]
\begin{tabular}{ll||cc|ccc|ccc}
&&\multicolumn{2}{c|}{Transformation matrix $W^1$} & \multicolumn{3}{c|}
{Conversion matrix $A^1$ on private resource}& \multicolumn{3}{c}{Conversion matrix $A^1$ on Commons}\\
     Var&Activity&Health&Pleasure&Money&Time&Car & Road $ij$ & PT $ij$  & Park\\\hline \hline
     $x_{ijr}$&Take the road $ij$&0&-1&2&0.5&0.5& $\epsilon$ & 0&0\\
     $x_{ijc}$&Take the pubic transport $ij$ &0&-2&2&0.5&0.5 & 0 & $\epsilon$&0\\\hline
      $y_{11}$  & Sleep &10&10&0&9&0&0&0&0\\
     $y_{12}$&Family Time &0&2&0&1&0&0&0&0\\\hline
     $y_{21}$ & Walk &3&2&0&2&0&0&0&$\epsilon$\\
     $y_{22}$& Run &6&-2&0&1&0&0&0&$\epsilon$\\\hline
     $y_{31}$& Museum &1&6&10&3&0&0&0&0\\
     $y_{32}$& After work &-1&1&5&1&0&0&0&0\\
     $y_{33}$& Doctor &5&-1&30&1&0&0&0&0\\\hline
     $y_{41}$&Work&1&1&-100&8&0&0&0&0\\
\end{tabular}
    \caption{Table of the conversion and transformation matrix of citizen $1$}
    \label{fig:tab1}
\end{table}

\begin{table}[H]

\begin{tabular}{ll||cc|ccc|ccc}
&&\multicolumn{2}{c|}{Transformation Matrix $W^2$} & \multicolumn{3}{c|}
{Conversion matrix $A^2$ on private resources}& \multicolumn{3}{c}{Conversion matrix $A^2$ on Commons}\\
     Var&Activity&Health&Pleasure&Money&Time&Car & Road $ij$ & PT $ij$ & Park  \\\hline \hline
     $x_{ijr}$&Take the road $ij$&0&0&2&0.5&0.5& $\epsilon$ & 0& 0\\
     $x_{ijc}$&Take the pubic transport $ij$ &0&-1&2&0.5&0.5 & 0 &$\epsilon$ & 0\\\hline
      $y_{11}$  & Sleep &10&10&0&9&0&0&0& 0\\
     $y_{12}$&Family Time &0&1&0&1&0&0&0&0\\\hline
     $y_{21}$ & Walk &1&2&0&2&0&0&0&$\epsilon$\\
     $y_{22}$& Run &3&4&0&1&0&0&0&$\epsilon$\\\hline
     $y_{31}$& Museum &0&2&10&3&0&0&0&0\\
     $y_{32}$& After work &0&2&5&1&0&0&0&0\\
     $y_{33}$& Doctor &2&-3&30&1&0&0&0&0\\\hline
     $y_{41}$&Work&1&1&-100&8&0&0&0&0\\
\end{tabular}
    \caption{Table of the conversion and transformation matrix of citizen $2$}
    \label{fig:tab2}
\end{table}
\end{landscape}
\begin{table}
    \centering

\begin{tabular}{c|ccc}
     Agent & Money & Time in hours & Car in hours  \\\hline
     $1$ & 0 & 24 & 24\\
     $2$ & 0 & 24 & 0\\
\end{tabular}
\caption{Vector of private resources ($R^i$) of citizens $1$ and $2$}
    \label{fig:commodities}
\end{table}

\begin{figure}
    \centering
\normalsize
\begin{tikzpicture}
\begin{axis}[
    title={},
    xlabel={Health},
    ylabel={Pleasure},
    xmin=0, xmax=28,
    ymin=0, ymax=44,
    xtick={0,4,8,12,16,20,24},
    ytick={0,4,8,12,16,20,24, 28,32,36,40},
    grid=major,
    axis x line=bottom, axis y line = left,
    %legend pos=north west,
    %ymajorgrids=true,
    %grid style=dashed,
]
%\node [above right, color=red] at (100,400) {$s_1$};
%\node [above right, color=red] at (110,210) {$s_2$};
%\node [above right, color=red] at (120,180) {$s_3$};
%\node [above right, color=red] at (160,150) {$s_4$};
%\node [above right, color=red] at (170,130) {$s_5$};
%\node [above right, color=red] at (200,120) {$s_6$};
%\node [above right, color=red] at (220,100) {$s_7$};
%\node [above right, color=red] at (250,80) {$s_8$};

%\node [below left, color=green] at (100,250) {$s_1$};
%\node [below left, color=green] at (110,170) {$s_2$};
%\node [below left, color=green] at (140,150) {$s_3$};
%\node [below left, color=green] at (150,120) {$s_4$};
%\node [below left, color=green] at (160,100) {$s_5$};
%\node [below left, color=green] at (170,70) {$s_6$};

\addplot[color= red, mark=square,
]
coordinates {(11,21)};
\addplot[color= red, mark=square,
]
coordinates {(10,40)};
\addplot[color= red, mark=square,
]
coordinates {(11,21)};
\addplot[color= red, mark=square,
]
coordinates {(20,12)};
\addplot[color= red, mark=square,
]
coordinates {(22,10)};
\addplot[color= red, mark=square,
]
coordinates {(17,13)};
\addplot[color= red, mark=square,
]
coordinates {(16,15)};
\addplot[color= red, mark=square,
]
coordinates {(25,8)};
\addplot[color= red, mark=square,
]
coordinates {(12,18)};
\addplot[color= red, mark=square,
]
coordinates {(25, 8)};

\addplot[color= green, mark=square,
]
coordinates {(10,25)};
\addplot[color= green, mark=square,
]
coordinates {(11,17)};
\addplot[color= green, mark=square,
]
coordinates {(14,15)};
\addplot[color= green, mark=square,
]
coordinates {(15,12)};
\addplot[color= green, mark=square,
]
coordinates {(16,10)};
\addplot[color= green, mark=square,
]
coordinates {(17,8)};

\addplot[color= blue, mark=square,
]
coordinates {(6,14)};
\addplot[color= blue, mark=square,
]
coordinates {(7,15)};
\addplot[color= blue, mark=square,
]
coordinates {(8,16)};
\addplot[color= blue, mark=square,
]
coordinates {(9,17)};
\addplot[color= blue, mark=square,
]
coordinates {(10,18)};

\end{axis}
\end{tikzpicture}
    \caption{Citizens Welfare representation in the initial situation. In red
    the Welfare representation of 1 $\mathcal{X}_1: $ $\mathcal{W}(\mathcal{X}_1)$.
    In blue, some solutions that are in the Beings set but not the Welfare representation of 1.
    In green the Welfare representation of 2 $\mathcal{X}_2:$ $W(\mathcal{X}_2)$}.
    \label{fig:NORMAL}
\end{figure}
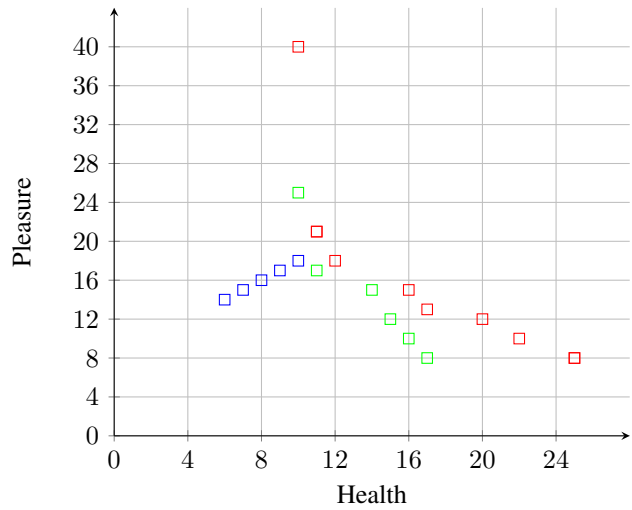

The Welfare representations $\mathcal{W}(\mathcal{X}_1)$ and
$\mathcal{W}(\mathcal{X}_2)$ are graphically represented in Figure
\ref{fig:NORMAL}. To obtain these solutions, we solved two linear programs;
Table \ref{pl} in Annex 1 displays the linear program corresponding to
$\mathcal{W}(\mathcal{X}_1$).

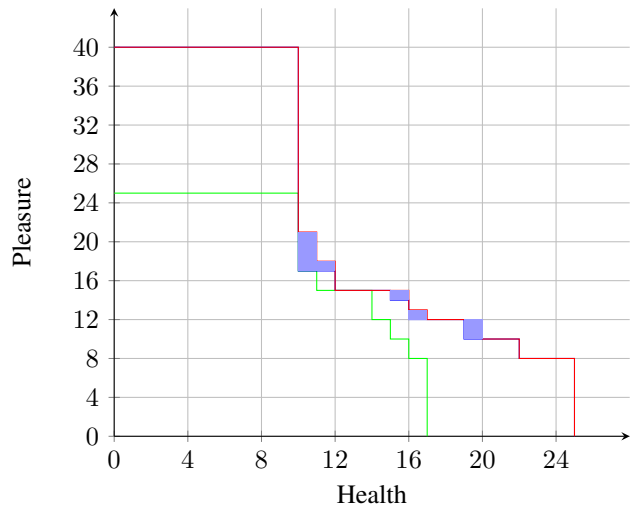
\begin{figure}
    \centering
\begin{tikzpicture}
\begin{axis}[
    title={},
    xlabel={Health},
    ylabel={Pleasure},
    xmin=0, xmax=28,
    ymin=0, ymax=44,
    xtick={0,4,8,12,16,20,24},
    ytick={0,4,8,12,16,20,24, 28,32,36,40},
    grid=major,
    axis x line=bottom, axis y line = left,
    %legend pos=north west,
    %ymajorgrids=true,
    %grid style=dashed,
]

\addplot[color= green,
]
coordinates {(0,25)(10,25)(10,17)(11,17)(11,15)(14,15)(14,12)(15,12)(15,10)(16,10)(16,8)(17,8)(17,0)};
\addplot[color= blue,
]
coordinates {(0,40)(10,40)(10,17)(12,17)(12,15)(15,15)(15,14)(16,14)(16,12)(19,12)(19,10)(22,10)(22,8)(25,8)(25,0)};
\addplot[color= red,
]
coordinates {(0,40)(10,40)(10,21)(11,21)(11,18)(12,18)(12,15)(16,15)(16,13)(17,13)(17,12)(20,12)(20,10)(22,10)(22,8)(25,8)(25,0)};

\fill[blue!40!white] (100,170) rectangle (110,210);
\fill[blue!40!white] (110,170) rectangle (120,180);
\fill[blue!40!white] (150,140) rectangle (160,150);
\fill[blue!40!white] (160,120) rectangle (170,130);
\fill[blue!40!white] (190,100) rectangle (200,120);

\end{axis}
\end{tikzpicture}
    \caption{$W(\mathcal{X}_1) =$ the Pareto-frontier in red,
    $W(\mathcal{X}_1 \backslash Road_{14}) =$ the Pareto-frontier in red and blue,
    $W(\mathcal{X}_2 \backslash Road_{14}) =W(\mathcal{X}_2)= $ the Pareto-frontier in green.}
    \label{fig:sansroute}
\end{figure}

Then, it is possible to represent how the welfare of citizens change in
different scenarios, after the damage suffered by a Common, by modifying the
parameters associated with a Common from $1$ to $0$ (for the sake of simplicity we will not change other parameters). We solve the new LP; for instance Table \ref{plsanspark} in Annex 1 displays the LP used in order to obtain the new welfare of citizen $1$ after the damage of the park.  Figure
\ref{fig:sansroute} represents the welfare of 1: $W(\mathcal{X}_1 | Road_{14})$ and 2: $W(\mathcal{X}_2 | Road_{14})$ after the damage of the road between their 1:home and their 4:work place. Figure \ref{fig:sanspark} represents the welfare of 1: $W(\mathcal{X}_1 | park)$ and 2: $W(\mathcal{X}_2 | park)$ after the damage of the park.

These two figures allow a simple analysis of the negative impact of the
damage of a Common. For instance, the damage to the Road linking 1 to 4 does not have a high negative impact on citizens' Welfare representation. Indeed, it does not affect the Welfare representation of 2, and the possible Beings of 1 are ``decreased'' but are still close to what they were. On the other hand, the damage to the park can be seen as having a high impact, specially because of its impact on health. The beings with high Health are not achievable anymore.

\normalsize

%\begin{tabular}{ccccccccccccccccc}
%     sol & y_{11} & y_{12} & y_{21} & y_{22} & y_{31} & y_{32} & y_{33}
%& y_{41} & x_{23c} & x_{34c} & x_{12r} & x_{13r} & x_{14r} & x_{23r} & x_{24r} &x_{34r}    \\
%     A & 1 & 6 & 0 & 0 & 0& 0& 0&  1& 0& 0& 0&0&2&0&0&0\\
%     B & 1 & 4 & 0 & 0 & 0 & 0 & 1 & 1 & 0 & 0 & 0 & 1 & 1 &0 & 0 &1 \\
%     C & 1 & 1 & 0 & 0 & 1 & 0 & 1 & 1 & 0 & 0 &0 &1&1 &0 &0& 1\\
%     D & 1 & 2 & 1 & 1 &0 & 0 & 0 & 1 & 0 & 0 & 1 & 0 &1 &0 &1 & 0 \\
%     E & 1 & 3 &0 & 1 & 0&0 & 1 & 1 &0&0& 1 &0 & 1 &1 &0 & 1\\
%     F & 1 & 1 & 1 & 1 & 0 & 0 & 1 & 1 & 0 & 0 & 1 & 0 & 1 & 1 & 0 & 1\\
%\end{tabular}

\normalsize

\begin{figure}
    \centering
\begin{tikzpicture}
\begin{axis}[
    title={},
    xlabel={Health},
    ylabel={Pleasure},
    xmin=0, xmax=28,
    ymin=0, ymax=44,
    xtick={0,4,8,12,16,20,24},
    ytick={0,4,8,12,16,20,24, 28,32,36,40},
    grid=major,
    axis x line=bottom, axis y line = left,
    %legend pos=north west,
    %ymajorgrids=true,
    %grid style=dashed,
]
%\node [above right, color=red] at (100,400) {$s_1$};
%\node [above right, color=red] at (110,210) {$s_2$};
%\node [above right, color=red] at (120,180) {$s_mus$};
%\node [above right, color=red] at (160,150) {$s_3$};
%\node [above right, color=red] at (170,130) {$s_4$};
%\node [above right, color=red] at (200,120) {$s_5$};
%\node [above right, color=red] at (220,100) {$s_6$};
%\node [above right, color=red] at (250,80) {$s_7$};

%\node [below left, color=blue] at (100,400) {$s_1$};
%\node [below left, color=blue] at (120,170) {$s_2$};
%\node [below left, color=blue] at (150,150) {$s_3$};
%\node [below left, color=blue] at (160,140) {$s_4$};
%\node [below left, color=blue] at (170,130) {$s_5$};
%\node [below left, color=blue] at (190,120) {$s_6$};
%\node [below left, color=blue] at (200,110) {$s_7$};
%\node [below left, color=blue] at (220,100) {$s_8$};
%\node [below left, color=blue] at (250,80) {$s_9$};
%\fill[blue!40!white] (0,0) rectangle (40,40);

%\fill[blue!40!white] (0,0) rectangle (40,40);
\fill[blue!40!white] (220,0) rectangle (250,80);
\fill[blue!40!white] (200,0) rectangle (220,100);
\fill[blue!40!white] (170,0) rectangle (200,120);

\fill[pink!40!white] (160,0) rectangle (170,70);
\fill[pink!40!white] (150,0) rectangle (160,100);
\fill[pink!40!white] (140,0) rectangle (150,120);
\fill[pink!40!white] (130,0) rectangle (140,150);
\fill[pink!40!white] (110,110) rectangle (130,150);
\addplot[color= red, mark=square,
]
coordinates {(11,21)};
\addplot[color= red, mark=square,
]
coordinates {(10,40)};
\addplot[color= red, mark=square,
]
coordinates {(17,13)};
\addplot[color= red, mark=square,
]
coordinates {(16,15)};
\addplot[color= red, mark=square,
]
coordinates {(12,18)};

\addplot[color= green, mark=square,
]
coordinates {(10,25)};
\addplot[color= green, mark=square,
]
coordinates {(11,17)};
\addplot[color= green, mark=square,
]
coordinates {(13,11)};
\addplot[color= blue,
]
coordinates {(0,40)(10,40)(10,21)(11,21)(11,18)(12,18)(12,15)(16,15)(16,13)(17,13)(17,12)(20,12)(20,10)(22,10)(22,8)(25,8)(25,0)};

\addplot[color= pink,
]
coordinates {(0,25)(10,25)(10,17)(11,17)(11,15)(14,15)(14,12)(15,12)(15,10)(16,10)(16,7)(17,7)(17,0)};

\addplot[color= green,
]
coordinates {(11,17)(11,11)(13,11)(13,0)};

\addplot[color= red,
]
coordinates {(17,0)(17, 13)};

\end{axis}

\end{tikzpicture}
\caption{$W(\mathcal{X}_1 \backslash Park) = $ set of points red,
    $W(\mathcal{X}_2\backslash Park) = $ set of points in green}
\label{fig:sanspark}
\end{figure}

\section*{Conclusion}

A common rational argument justifying several public policies or their designs is that if and when implemented (these policies) the welfare of the citizens will improve or will be protected from potential threats. However, little is specifically said as far as the measurement of welfare is concerned in order to be able to check whether such ``predictions'' (or promises) are well founded and sound. Welfare is measured either through simple figures (such as the GDP) or when specific projects are to be assessed welfare is practically considered as the sum of the utilities of undistinguished consumers (using Cost-Benefit Analysis). Such approaches are theoretically poor and practically of little help although allow to establish a rational justification. The aim of this paper is to construct an alternative framework allowing to construct sound assessments (about the modification of welfare) which should also be of some practical interest.

For this purpose we first do a rather extensive survey of how welfare has been considered within different approaches and we focus specifically to the so called Capability Approach introduced by A. Sen in the 80s. The advantage of this approach consists in considering explicitly the subjective difference of how welfare is constructed and perceived by the citizens through their acts and beings and not just because of their endowments. However, the capability approach, despite opening interesting theoretical opportunities and despite some fine applications it is far from being a really operational tool which can be generally used in order to model and assess welfare for some decisions making and aiding purposed.

For this reason we suggest a mathematical programming formulation of a single citizen's welfare given his/her private endowments, access to a set of ``Commons'' and considering a set of potential actions and different welfare dimensions considered interesting (by that citizen).

This suggestion opens two interesting directions. On the one hand we can cluster citizens not on the basis of endowments or socio-demographic characteristics but in the basis of the similarity of their capability sets (these being the Pareto frontiers of their welfare). On the other hand we can simulate different scenarios where private endowments and/or the Commons can be damaged and different policies aimed at ``protecting them''. Such simulations should provide fundamental rational insight as far as the design of public policies is concerned.

\bibliographystyle{apalike}
\bibliography{cahier05}
%\bibliography{cahier02.bbl}

\newpage
    \begin{landscape}% Landscape page
        \centering % Center table
\section*{Annex 1}
\begin{table}[H]
\centering
\resizebox{1.6\textwidth}{!}{
\begin{tabular}{lllllllllllll}
    $\max$&$($  &  &$10y_{11}$ & &$+3y_{21}$ &$+6y_{22}$ &$+y_{31}$ & $-y_{32}$ &$5y_{33}$ &$+y_{41})$ \\
     $\max$ & $(-\sum\limits_{ijr\in E} x_{ijr}$  & $-\sum\limits_{ijc\in E} 2x_{ijc}$
     &$+10y_{11}$ & $+2y_{12}$ &$+2y_{21}$ &$-2y_{22}$ &$+6y_{31}$ & $+y_{32}$ &$-2y_{33}$ &$+y_{41})$ \\\\
     S.T.& $2\sum\limits_{ijr\in E} x_{ijr}$ & $+2\sum\limits_{ijc\in E} x_{ijc} $
     & & & & & $+10y_{31}$ & $+5y_{32}$ & $+30y_{33}$ & $-100y_{41}$ & $\leq$ & 0\\
     & $\frac{1}{2}\sum\limits_{ijr\in E} x_{ijr}$  & $+\frac{1}{2}\sum\limits_{ijc\in E} 2x_{ijc}$
     &$+9y_{11}$ & $+y_{12}$ &$+2y_{21}$ &$+2y_{22}$ &$+3y_{31}$ & $+y_{32}$ &$+y_{33}$ &$+8y_{41}$ & $\leq$ & 24 \\
     &$\frac{1}{2}\sum\limits_{ijr\in E} x_{ijr}$  & $+\frac{1}{2}\sum\limits_{ijc\in E} 2x_{ijc}$ & & & & & &  & && $\leq$ & 24 \\
     & $\epsilon x_{ijr}$  &  & &  & & & &  &&&$\leq$ & 1 $\forall ijr \in E$  \\
     &   & $\epsilon x_{ijc}$ & &  & & & &  &&&$\leq$ & 1 $\forall ijc \in E$\\
     &  &  &$y_{11}$&$+y_{12}$ & & & &&  & & $\leq$ &$M(\sum\limits_{i1r\in E} + \sum\limits_{i1c\in E})$\\
     &  &  && &$y_{21}$ &$+y_{22}$ & &&  & & $\leq$ &$M(\sum\limits_{i2r\in E} x_{i2r} + \sum\limits_{i2c\in E} x_{i2c})$\\
     &  &  & & & & & $y_{31}$ & $+y_{32}$ & $+y_{33}$ & & $\leq$ &$M(\sum\limits_{i3r\in E}  x_{i3r} + \sum\limits_{i3c\in E} x_{i3c})$\\
     &  &  && & & & &&  & $+y_{41}$ & $\leq$ &$M(\sum\limits_{i4r\in E}  x_{i4r} + \sum\limits_{i4c\in E}  x_{i4c})$\\
     & $\sum\limits_{i1r\in E} x_{i1r}-\sum\limits_{1jr\in E}x_{1jr}$ &$+\sum\limits_{i1c\in E}x_{i1c}-\sum\limits_{1jr\in E}x_{1jc}$ &&&&&&&&&$=$&$0$\\
     & $\sum\limits_{i2r\in E} x_{i2r}-\sum\limits_{2jr\in E}x_{2jr}$ &$+\sum\limits_{i2c\in E}x_{i2c}-\sum\limits_{2jr\in E}x_{2jc}$ &&&&&&&&&$=$&$0$\\
     & $\sum\limits_{i3r\in E} x_{i3r}-\sum\limits_{3jr\in E}x_{3jr}$ &$+\sum\limits_{i3c\in E}x_{i3c}-\sum\limits_{3jr\in E}x_{3jc}$ &&&&&&&&&$=$&$0$\\
     & $\sum\limits_{i4r\in E} x_{i4r}-\sum\limits_{4jr\in E}x_{4jr}$ &$+\sum\limits_{i4c\in E}x_{i4c}-\sum\limits_{4jr\in E}x_{4jc}$ &&&&&&&&&$=x$&$0$\\
     &$x_{ijr},$ & $x_{ijc}$ &&&&&&&&&$\in$& $\mathcal{N}, \forall ijr, ijc \in E$ \\
     &&&$y_{11},$& & $y_{21},$ & $y_{22},$ & $y_{31},$ & & $y_{33},$ & $y_{41}$ & $\in$ & $\{0,1\}$\\
     &&&&$y_{12},$ & &  & & $y_{32}$ & & & $\in$ & $\mathcal{N}$ \\
\end{tabular}}
\caption{ILP for $\mathcal{X}_1$}
\label{pl}
\end{table}
\end{landscape}
\begin{landscape}% Landscape page
\centering % Center table
\begin{table}[]
\centering
\resizebox{1.6\textwidth}{!}{
\begin{tabular}{lllllllllllll}
    $\max$&$($  &  &$10y_{11}$ & &$+3y_{21}$ &$+6y_{22}$ &$+y_{31}$ & $-y_{32}$ &$5y_{33}$ &$+y_{41})$ \\
     $\max$ & $(-\sum\limits_{ijr\in E} x_{ijr}$  & $-\sum\limits_{ijc\in E} 2x_{ijc}$
     &$+10y_{11}$ & $+2y_{12}$ &$+2y_{21}$ &$-2y_{22}$ &$+6y_{31}$ & $+y_{32}$ &$-2y_{33}$ &$+y_{41})$ \\\\
     S.T.& $2\sum\limits_{ijr\in E} x_{ijr}$ & $+2\sum\limits_{ijc\in E} x_{ijc} $
     & & & & & $+10y_{31}$ & $+5y_{32}$ & $+30y_{33}$ & $-100y_{41}$ & $\leq$ & 0\\
     & $\frac{1}{2}\sum\limits_{ijr\in E} x_{ijr}$  & $+\frac{1}{2}\sum\limits_{ijc\in E} 2x_{ijc}$
     &$+9y_{11}$ & $+y_{12}$ &$+2y_{21}$ &$+2y_{22}$ &$+3y_{31}$ & $+y_{32}$ &$+y_{33}$ &$+8y_{41}$ & $\leq$ & 24 \\
     &$\frac{1}{2}\sum\limits_{ijr\in E} x_{ijr}$  & $+\frac{1}{2}\sum\limits_{ijc\in E} 2x_{ijc}$ & & & & & &  & && $\leq$ & 24 \\
     & $\epsilon x_{ijr}$  &  & &  & & & &  &&&$\leq$ & 1 $\forall ijr \in E$  \\
     &   & $\epsilon x_{ijc}$ & &  & & & &  &&&$\leq$ & 1 $\forall ijc \in E$\\
     & & & &  & {\color{red}$\epsilon y_{21}$} &{\color{red}$ + \epsilon y_{22}$} &&  &&&{\color{red}$\leq$ }& {\color{red}0 }\\
     &  &  &$y_{11}$&$+y_{12}$ & & & &&  & & $\leq$ &$M(\sum\limits_{i1r\in E} + \sum\limits_{i1c\in E})$\\
     &  &  && &$y_{21}$ &$+y_{22}$ & &&  & & $\leq$ &$M(\sum\limits_{i2r\in E} x_{i2r} + \sum\limits_{i2c\in E} x_{i2c})$\\
     &  &  & & & & & $y_{31}$ & $+y_{32}$ & $+y_{33}$ & & $\leq$ &$M(\sum\limits_{i3r\in E}  x_{i3r} + \sum\limits_{i3c\in E} x_{i3c})$\\
     &  &  && & & & &&  & $+y_{41}$ & $\leq$ &$M(\sum\limits_{i4r\in E}  x_{i4r} + \sum\limits_{i4c\in E}  x_{i4c})$\\
     & $\sum\limits_{i1r\in E} x_{i1r}-\sum\limits_{1jr\in E}x_{1jr}$ &$+\sum\limits_{i1c\in E}x_{i1c}-\sum\limits_{1jr\in E}x_{1jc}$ &&&&&&&&&$=$&$0$\\
     & $\sum\limits_{i2r\in E} x_{i2r}-\sum\limits_{2jr\in E}x_{2jr}$ &$+\sum\limits_{i2c\in E}x_{i2c}-\sum\limits_{2jr\in E}x_{2jc}$ &&&&&&&&&$=$&$0$\\
     & $\sum\limits_{i3r\in E} x_{i3r}-\sum\limits_{3jr\in E}x_{3jr}$ &$+\sum\limits_{i3c\in E}x_{i3c}-\sum\limits_{3jr\in E}x_{3jc}$ &&&&&&&&&$=$&$0$\\
     & $\sum\limits_{i4r\in E} x_{i4r}-\sum\limits_{4jr\in E}x_{4jr}$ &$+\sum\limits_{i4c\in E}x_{i4c}-\sum\limits_{4jr\in E}x_{4jc}$ &&&&&&&&&$=$&$0$\\
     &$x_{ijr},$ & $x_{ijc}$ &&&&&&&&&$\in$& $\mathcal{N}, \forall ijr, ijc \in E$ \\
     &&&$y_{11},$& & $y_{21},$ & $y_{22},$ & $y_{31},$ & & $y_{33},$ & $y_{41}$ & $\in$ & $\{0,1\}$\\
     &&&&$y_{12},$ & &  & & $y_{32}$ & & & $\in$ & $\mathcal{N}$
\end{tabular}
}
\caption{ILP for $\mathcal{X}_1 \backslash park$}
\label{plsanspark}
\end{table}
    \end{landscape}

\end{document}